\documentclass{cernyrep}
\usepackage{graphics}
\usepackage{color}
\usepackage{amsmath,amsthm,amsfonts,amssymb,multirow}

\newcommand\smfrac[2]{{\textstyle{\frac{#1}{#2}}}}
\newcommand\as{\alpha_{\scriptscriptstyle\mathrm{S}}}
\newcommand\slsh{{}\!\!\not{}\!}
\newcommand\ldot{\!\cdot\!}
\newcommand\Tr{\mathrm{Tr}}

\newcommand\be{\begin{equation}}
\newcommand\ee{\end{equation}}
\newcommand\bee{\begin{eqnarray}}
\newcommand\eee{\end{eqnarray}}

\renewcommand{\thesection}{\arabic{section}}
\renewcommand{\thesubsection}{\arabic{section}.\arabic{subsection}}
\renewcommand{\thesubsubsection}{\arabic{section}.\arabic{subsection}.\arabic{subsubsection}}

\begin{document}

\pagestyle{empty}


\begin{titlepage}
\begin{flushright}
\begin{tabular}{r}
CERN-PH-TH/2009-194\\
MAN/HEP/2009/35
\end{tabular}
\end{flushright}

\vspace*{1.9truecm}

\begin{center}
{\LARGE \bf Quantum ChromoDynamics}

\vspace*{2.6cm}

\smallskip
\begin{center}
{\sc {\large Michael H. Seymour}}\\
\vspace*{2mm}
{\sl {\large School of Physics and Astronomy, University of Manchester,
    Manchester, M13 9PL, U.K., and\\
    Theoretical Physics Group, CERN,
    CH-1211 Geneva 23, Switzerland}}
\end{center}

\vspace{2.6truecm}

{\large\bf Abstract\\[10pt]} \parbox[t]{\textwidth}{
These lectures on QCD stress the theoretical elements that underlie a
wide range of phenomenological studies, particularly gauge invariance,
renormalization, factorization and infrared safety.  The three parts
cover the basics of QCD, QCD at tree level and higher order
corrections.}

\vspace{2.6cm}

{\sl
Lectures given at the {\it 2009 Latin American School of High Energy Physics,}\\
Recinto Quirama, Antioquia, Colombia, 15 -- 28 March 2009,\\
pp. 97--143 of the proceedings, CERN--2010-001.
}
\end{center}

\end{titlepage}

\thispagestyle{empty}
\vbox{}
\newpage


\pagestyle{plain}

\setcounter{page}{1}
\pagenumbering{roman}

\tableofcontents

\newpage

\setcounter{page}{1}
\pagenumbering{arabic}

\title{Quantum ChromoDynamics}
\author{Michael H. Seymour}
\institute{University of Manchester, UK, and CERN, Geneva, Switzerland}
\maketitle
\begin{abstract}
These lectures on QCD stress the theoretical elements that underlie a
wide range of phenomenological studies, particularly gauge invariance,
renormalization, factorization and infrared safety.  The three parts
cover the basics of QCD, QCD at tree level, and higher order corrections.
\end{abstract}

\section{Basics of QCD}
\subsection{Introduction}

QCD is the theory of the strong nuclear force, one of the four
fundamental forces of nature.  It describes the interactions of quarks,
via their colour quantum numbers.  It is an unbroken gauge theory.  The
gauge bosons are gluons.  It has a similar structure to QED, but with
one important difference: the gauge group is non-Abelian, SU(3), and
hence the gluons are self-interacting.  This results in a negative
$\beta$-function and hence asymptotic freedom at high energies and
strong interactions at low energies.

These strong interactions are confining: only colour-singlet states can
propagate over macroscopic distances.  The only stable colour singlets
are quark--antiquark pairs, mesons, and three-quark states, baryons.  In
high energy reactions, like deep inelastic scattering, the quark and
gluon constituents of hadrons act as quasi-free particles, partons.
Such reactions can be factorized into the convolution of
non-perturbative functions that describe the distribution of partons in
the hadron, which cannot be calculated from first principles (at
present) but are universal (process-independent), with process-dependent
functions, which can be calculated as perturbative expansions in the
coupling constant $\as$.

Beyond leading order in $\as$, the parton distribution functions and
coefficient functions become intermixed.  They can still be factorized,
but the parton distribution functions become energy-dependent.  Although
the input distributions at some fixed energy scale still cannot be
calculated, the energy dependence is given by perturbative evolution
equations.

In sufficiently inclusive cross sections, called infrared safe, the
non-perturbative distributions cancel and distributions can directly be
calculated in perturbation theory.  Non-perturbative corrections are
then suppressed by powers of the high energy scale.  The most important
examples are jet cross sections, where jets of hadrons have a direct
connection to the perturbatively-calculable quarks and gluons.

This course will attempt to give a brief overview of the subject.  The
approach will be pretty phenomenological, with most results stated
rather than derived.  I will however attempt to sketch in most cases
roughly how they would be derived.  One thing I will not have time to go
into in much detail will be heavy quarks: in most cases we will treat
the d, u, s, c and b quarks as massless and neglect the top quark, an
approximation that I will motivate in Section~\ref{sec:decoupling}.

It is hard to give a better introduction to the subject than the book
`QCD and Collider Physics', by Keith Ellis, James Stirling and Bryan
Webber~\cite{ESW}.  So I will follow the ESW approach and notation pretty
closely.  In most cases they will be able to give you a few more details
and references to much more detailed treatments if you want to go
further.  For a much more detailed treatment of the formulation of QCD
and its renormalization in particular Peskin and Schroeder~\cite{Peskin}
is also unbeatable.

As there are many parallels with QED I will have to assume prior
knowledge of the basics of QED and that you can calculate a few simple
cross sections.  However we start by recapitulating a few features.

\subsection{Basics of QED}

QED is a gauge theory with gauge group U(1).  It can be derived using
the gauge principle.  The classical Lagrangian density for $n$ types of
non-interacting fermion is
\be
  \label{LQEDferm}
  {\cal L}_{\mathrm{ferm}} = \sum_i^n \bar{f}_i(i\slsh\partial-m_i)f_i,
\ee
where $f_i$ is a spinor-valued wave function describing plane waves of
momentum $p_i$, $\bar{f}_i$ its Dirac conjugate $f_i^\dagger\gamma^0$,
$\slsh a$ is shorthand for $\gamma^\mu a_\mu$ and $\gamma^\mu$ are Dirac
spinor matrices with anticommutation relation
\be
  \{\gamma^\mu,\gamma^\nu\} = 2g^{\mu\nu}.
\ee
The Lagrangian density~(\ref{LQEDferm}) is invariant under global
changes of gauge,
\be
  f_i \to f'_i=\exp(ie_i\theta)f_i,
\ee
where $e_i$ is an arbitrary flavour-dependent parameter, which will turn
out to be proportional to electric charge.  We can derive QED by asking
how we would need to modify~(\ref{LQEDferm}) to make it also invariant
under local changes of gauge,
\be
  \label{QEDlocal}
  f_i(x) \to f'_i(x)=\exp(ie_i\theta(x))f_i(x).
\ee
This can be done by introducing a new vector-valued field $A_\mu$, which
transforms under the same change of gauge like
\be
  A_\mu(x) \to A'_\mu(x)=A_\mu(x)
    +\frac{i}{e}\Bigl(\partial_\mu\exp(i\theta(x))\Bigr)\exp(-i\theta(x)),
\ee
and by replacing the derivative $\partial_\mu$ by the covariant
derivative,
\be
  D_\mu = \partial_\mu + ie\hat Q\,A_\mu,
\ee
where $\hat Q$ is the charge operator, defined by
\be
  \hat Q\,f_i = e_if_i.
\ee

Since $A_\mu$ is a new field that we have introduced, we must make it
physical by adding a kinetic term,
\be
  {\cal L}_{\mathrm{kin}} = -\frac14 F_{\mu\nu}F^{\mu\nu},
\ee
where the field strength tensor $F_{\mu\nu}$ is defined by
\be
  \label{QEDFmunu}
  F_{\mu\nu} = \partial_\mu A_\nu - \partial_\nu A_\mu.
\ee
The classical QED Lagrangian density is therefore given by
\be
  \label{QEDclass}
  {\cal L}_{\mathrm{classical}} = -\frac14 F_{\mu\nu}F^{\mu\nu}
  +\sum_i^n \bar{f}_i(i\slsh D-m_i)f_i.
\ee
This is now invariant under local changes of gauge.

Perturbative calculations are made according to the Feynman rules.
These can be read off from the action, defined by
\be
  S = i\int d^4x \, {\cal L}.
\ee
There is however one complication.  The photon propagator
$\Delta_{\gamma,\mu\nu}(p)$ is derived from the inverse of the bilinear
term in $A_\mu$:
\be
  \Delta_{\gamma,\mu\nu}(p) \times
  i\Bigl[p^2g^{\nu\sigma} - p^\nu p^\sigma\Bigr] = \delta_\mu^\sigma.
\ee
This does not have an inverse.  However, we can exploit the gauge
invariance of the theory to rewrite it in a physically equivalent form
that is invertible.  Since the Lagrangian density is gauge invariant, we
can choose some convenient gauge to work in and the final answer should
be independent of which we chose.  For example, in the covariant gauge,
we have the condition
\be
  \label{QEDconstraint}
  \partial^\mu A_\mu=0
\ee
at every space-time point.  We can therefore add an extra term to the
Lagrangian density
\be
  {\cal L}_{\mathrm{gauge-fixing}} =
  -\frac1{2\lambda}\left(\partial^\mu A_\mu\right)^2,
\ee
where $\lambda$ is an arbitrary parameter, and provided we work in a
covariant gauge we cannot have changed the physics, since we have only
added zero.  (This is essentially just the method of undetermined
Lagrange multipliers for minimizing an action subject to a constraint:
the constraint is~(\ref{QEDconstraint}) and the multiplier is
$1/2\lambda$.)  The final results must clearly be independent of
$\lambda$, although it will appear at intermediate steps of
calculations.  Common choices are $\lambda=1$ (Feynman gauge) and
$\lambda\to0$ (Landau gauge).  For arbitrary~$\lambda$, we must now
solve
\be
  \Delta_{\gamma,\mu\nu}(p) \times
  i\Bigl[p^2g^{\nu\sigma} - (1-\frac1\lambda)p^\nu p^\sigma\Bigr] =
  \delta_\mu^\sigma,
\ee
which yields
\be
  \Delta_{\gamma,\mu\nu} =
  \frac{i}{p^2}\left(-g_{\mu\nu}+(1-\lambda)\frac{p_\mu p_\nu}{p^2}\right).
\ee
Clearly the Feynman gauge offers significant calculational advantages,
so we use it for most of the rest of this course.

Another popular class of gauges are the axial (or physical) gauges,
defined in terms of an arbitrary vector $n$, by
\be
  {\cal L}_{\mathrm{gauge-fixing}} =
  -\frac1{2\lambda}\left(n^\mu A_\mu\right)^2.
\ee
These have the result that an on-shell photon has two polarization
states, which, in the \mbox{$(n\!+\!p)$} rest-frame, are purely
transverse to its direction.  The penalty is that the propagator becomes
more complicated,
\be
  \Delta_{\gamma,\mu\nu} =
  \frac{i}{p^2}\left(-g_{\mu\nu}
  +\frac{n_\mu p_\nu+p_\mu n_\nu}{n\ldot p}
  -\frac{(n^2+\lambda\,p^2)p_\mu p_\nu}{(n\ldot p)^2}\right).
\ee
Obviously some simplification is obtained by setting $n^2=0$ and
$\lambda\to0$ (the `lightcone' gauge), but practical calculations are
still considerably more complicated than in covariant gauges.  In
particular, if making a numerical calculation, it is difficult to
guarantee that the spurious singularities $n\ldot p\to0$ cancel as they
should.

We therefore have the Feynman rules (in Feynman gauge):
\bee
  \Delta_i &=& \frac{i}{\slsh p-m_i} = i\frac{\slsh p+m_i}{p^2-m_i^2}, \\
  \Delta_{\gamma,\mu\nu} &=& i\frac{-g_{\mu\nu}}{p^2}, \\
  \Gamma_{\gamma f_i\bar{f}_i}^\mu &=& -i\,e_ie\gamma^\mu.
\eee

To calculate the cross section for a given process, we must write down
all possible diagrams, use the Feynman rules to give us the amplitude
$i{\cal M}$, use Dirac algebra and trace theorems to calculate
$\sum|{\cal M}|^2$, where the sum is over all unobserved quantum
numbers for example spin, divide by the overcounting of incoming states,
and integrate over phase space:
\be
  \sigma = \frac{1}{S} \; \frac{1}{2s} \; \int d\Gamma \; \sum|{\cal M}|^2.
\ee
An element of $n$-body phase space is given by
\bee
  d\Gamma &=& \prod_{i=1}^n \left(
  \frac{d^4p_i}{(2\pi)^4}\;(2\pi)\delta(p_i^2-m_i^2)
  \right) (2\pi)^4\delta^4(p_{tot}-\textstyle{\sum_i^n p_i}) \\
  &=& \prod_{i=1}^n \left(
  \frac{d^3p_i}{(2\pi)^32E_i}
  \right) (2\pi)^4\delta^4(p_{tot}-\textstyle{\sum_i^n p_i}).
\eee

For example, the cross section for $e^+e^-\to\mu^+\mu^-$ is calculated
as follows.  The amplitude is
\bee
  i{\cal M} &=& \bar{v}(p_{e^+})(ie)\gamma^\mu u(p_{e^-})
  \; i\frac{-g_{\mu\nu}}{(p_{e^+}+p_{e^-})^2} \;
  \bar{u}(p_{\mu^-})(ie)\gamma^\nu v(p_{\mu^+}) \\
  &=& \frac{-ie^2}{(p_{e^+}+p_{e^-})^2}
  \bar{v}(p_{e^+})\gamma^\mu u(p_{e^-})
  \; \bar{u}(p_{\mu^-})\gamma_\mu v(p_{\mu^+})
\eee
and hence
\be
  \sum|{\cal M}|^2 = \frac{(4\pi\alpha)^2}{s^2}
  \;\Tr\left\{\slsh p_{e^+} \gamma^\mu \slsh p_{e^-} \gamma^\nu\right\}
  \;\Tr\left\{\slsh p_{\mu^-} \gamma_\mu \slsh p_{\mu^+} \gamma_\nu\right\},
\ee
where $\alpha=e^2/4\pi$ and $s=(p_{e^+}+p_{e^-})^2$, or
\bee
  \hspace*{-0.5cm}
  \sum|{\cal M}|^2 &=& \frac{16(4\pi\alpha)^2}{s^2}
  \;\left(p_{e^+}^\mu p_{e^-}^\nu+p_{e^-}^\mu p_{e^+}^\nu
    -p_{e^+}\ldot p_{e^-}g^{\mu\nu}\right)
  \;\left(p_{\mu^-\!\!,\mu} p_{\mu^+\!\!,\nu}
    +p_{\mu^+\!\!,\mu} p_{\mu^-\!\!,\nu}
    -p_{\mu^+}\ldot p_{\mu^-}g_{\mu\nu}\right)
  \hspace*{-0.5cm}
  \nonumber\\{}\\
  &=& 8(4\pi\alpha)^2\frac{t^2+u^2}{s^2},
\eee
where $t=(p_{e^-}-p_{\mu^-})^2$ and $u=(p_{e^-}-p_{\mu^+})^2=-s-t$.
The cross section is therefore
\bee
  \sigma &=& \frac{1}{4} \; \frac{1}{2s} \int_{-s}^0 \frac{dt}{8\pi s} \;
  8(4\pi\alpha)^2\frac{t^2+u^2}{s^2} \\
  &=& \frac{4\pi\alpha^2}{3s}.
\eee

\subsection{SU(3) and colour}

QCD can be derived in exactly the same way as QED: we start from the
Lagrangian density for a set of non-interacting quarks and modify it in
just such a way that it is invariant under changes of gauge.  The only
difference is that instead of the gauge transformation being a simple
phase (U(1) group), we consider a non-Abelian group SU($N_c$).  This has
several important consequences.  Fermion charges will come in $N_c$
different types, called colours, they will be quantized (in contrast to
the electric charges $e_i$, which could take any value) and, most
importantly, the gauge bosons will be self-interacting.

It has been well-known since the early days of QCD that there are three
colours, for example from baryon wave functions, the total $e^+e^-$
cross section (which is proportional to $N_c$) and $\pi^0$ decay rate
(which is proportional to $N_c^2$).  However, in most calculations it is
useful to keep the number of colours $N_c$ arbitrary until the very last
step when it is set equal to three.  The $N_c$-dependent coefficients
are a useful diagnostic tool in understanding the physical origins of
different terms, comparing different calculations and tracking down
errors.

We start by restating briefly some features of SU($N$), the group of $N\times
N$ unitary matrices ($U^\dagger U=1$) with determinant +1.  Let $U$ be
an element of SU($N$) that is infinitesimally close to the identity and
write it as
\be
  U = 1 + iG,
\ee
where $G$ has infinitesimal entries.  It must be hermitian
($G^\dagger=G$) and traceless.  One can choose a basis set of $N^2-1$
matrices, $t^A$, $A=1,\ldots,N^2-1$, such that any $G$ can be written as
\be
  G = \sum_A^{N^2-1} \epsilon^A t^A,
\ee
where $\epsilon_A$ are infinitesimal numbers.  Note that I will always
denote colour indices that run from 1 to $N$ by $a$ and from 1 to
$N^2\!-\!1$ by $A$.  The $t^A$ are called the generators of the group
and define its fundamental representation.  You can show that
$[t^A,t^B]$ is antihermitian and traceless and hence can be written as a
linear combination of other~$t^C$s,
\be
  \label{SU(N)algebra}
  [t^A,t^B] \equiv i\,f^{ABC}t^C,
\ee
where $f^{ABC}$ are a set of real constants, called the structure
constants of the group.  It is straightforward to see that $f^{ABC}$ is
antisymmetric in $A,B$, and with a little more work, one can prove that
it is antisymmetric in all its indices.  Equation~(\ref{SU(N)algebra})
defines the Lie algebra of the group.

We can also define a set of $(N^2\!-\!1)\times(N^2\!-\!1)$ matrices that
obey the same algebra:
\be
  \left(T^A\right)_{BC} \equiv -if^{ABC},
\ee
\be
  [T^A,T^B] =
  i\,f^{ABC}T^C.
\ee
These define the group's adjoint representation.

Although we started with elements infinitesimally close to the identity
matrix, we can calculate an arbitrary element $U$ by stringing together
an infinite number of infinitesimal elements,
\be
  U = \lim_{N\to\infty}(1+i\theta^At^A/N)^N = \exp(i\theta^At^A) \equiv
  \exp(it\ldot\theta).
\ee
Since $U$ is unitary and $t^A$ hermitian, we have
\be
  U^{-1} = \exp(-it\ldot\theta).
\ee

There are several identities we will require time and time again:
\bee
  \label{eq:colour1}
  \Tr(t^At^B) &=& \smfrac12 \delta^{AB} \equiv T_R\delta^{AB} \\
  \label{eq:colour2}
  \sum_A t_{ab}^At^A_{bc} &=& \frac{N^2-1}{2N} \delta_{ac} \equiv
  C_F\delta_{ac} \\
  \label{eq:colour3}
  \Tr(T^CT^D) &=& \sum_{A,B} f^{ABC}f^{ABD} = N \delta^{CD} \equiv
  C_A\delta^{CD},
\eee
where the constants $C_F$ and $C_A$ are the Casimir operators of the
fundamental and adjoint representations of the group respectively.
Although we know the numerical values of these constants:
\bee
  T_R &=& \frac12, \\
  C_F &=& \frac43, \\
  C_A &=& 3,\phantom{\frac12}
\eee
it is good practice, as I said, to leave them unexpanded in all
algebraic results.

In fact for practical calculations one only requires these, and other
similar, identities and never an explicit representation for $t^A$ or
$f^{ABC}$.

\subsection{The QCD Lagrangian}

The classical Lagrangian density for $n$ non-interacting quarks with
masses $m_i$ is
\be
  {\cal L}_{\mathrm{quarks}} = \sum_i^n
  \bar{q}^a_i(i\slsh\partial-m_i)_{ab}q^b_i,
\ee
where the factor $(i\slsh\partial-m_i)_{ab}$ is proportional to the
identity matrix in colour space.  This is invariant under global
SU($N_c$) transformations,
\be
  q_a \to q'_a = \exp(it\ldot\theta)_{ab}q_b.
\ee
To make it invariant under local transformations,
\be
  \label{eq:localSUN}
  q_a(x) \to q'_a(x) = \exp(it\ldot\theta(x))_{ab}q_b(x),
\ee
we have to introduce the covariant derivative,
\be
  D_{\mu,ab}=\partial_\mu\,1_{ab}+ig_s\,(t\ldot A_\mu)_{ab},
\ee
where $A^A_\mu$ are coloured vector fields that transform in just the
right way that we have
\be
  D'_{\mu,ab}q'_b(x)=\exp(it\ldot\theta(x))_{ab}D_{\mu,bc}q_c(x),
\ee
giving
\be
  \label{QCDAgauge}
  t\ldot A'_\mu = \exp(it\ldot\theta(x))\,t\ldot A_\mu\,\exp(-it\ldot\theta(x))
  +\frac{i}{g_s}\Bigl(\partial_\mu\exp(it\ldot\theta(x))\Bigr)
  \exp(-it\ldot\theta(x)).
\ee
We again have to introduce a kinetic term for this new field,
\be
  \label{QCDkin}
  {\cal L}_{\mathrm{kin}} = -\frac14 F_{\mu\nu}^AF^{\mu\nu}_A,
\ee
where $F_{\mu\nu}^A$ is the non-Abelian field strength tensor.  However,
the definition we used in QED~(\ref{QEDFmunu}) does not result in an
invariant Lagrangian density under transformation~(\ref{QCDAgauge}).
One must add an extra term,
\be
  F_{\mu\nu}^A = \partial_\mu A^A_\nu - \partial_\nu A^A_\mu
  -g_s f^{ABC}A_\mu^BA_\nu^C,
\ee
and only then is~(\ref{QCDkin}) invariant under gauge transformations.

This extra term has profound consequences for the theory: it means that
gluons are self-interacting, through three- and four-point vertices.
This will turn out to give rise to asymptotic freedom at high energies
and strong interactions at low energies, among the most fundamental
properties of QCD\@.  We therefore see that these are absolute
requirements of the SU($N_c$) gauge symmetry.

Before reading off the Feynman rules we again have to fix the gauge.
This proceeds in exactly the same way as in QED, leading to, in
covariant gauges,
\be
  {\cal L}_{\mathrm{gauge-fixing}} =
  -\frac1{2\lambda}\left(\partial^\mu A^A_\mu\right)^2.
\ee

Finally, it turns out that in a non-Abelian gauge theory, it is
necessary to add one extra term to the Lagrangian density, related to
the need for ghost particles.  These are beyond the scope of this
course, but basically they arise because when a non-Abelian gauge theory
is renormalized it is possible for unphysical degrees of freedom to
propagate freely.  These are cancelled off by introducing into the
theory an unphysical set of fields, the ghosts, which are scalars but
have Fermi statistics.  For practical purposes it is enough to know that
there exist Feynman rules for ghosts and that in every diagram with a
closed loop of internal gluons containing only triple-gluon vertices, we
must add a diagram with the gluons in the loop replaced
by ghosts.  It is worth noting that in physical gauges, as the name
suggests, ghost contributions always vanish and they can be ignored.

The final Lagrangian is therefore
\be
  {\cal L}_{\mathrm{QCD}} = -\frac14 F_{\mu\nu}^AF^{\mu\nu}_A
  +\sum_i^n \bar{q}^a_i(i\slsh D-m_i)_{ab}q^b_i
  -\frac1{2\lambda}\left(\partial^\mu A^A_\mu\right)^2
  +{\cal L}_{\mathrm{ghost}}.
\ee

\subsection{Feynman rules}

Just as in QED it is straightforward to read off the Feynman rules from
the action.  We obtain in Feynman gauge (only the gluon propagator is
gauge dependent)
\bee
  \Delta_i^{ab} &=& \delta^{ab}\frac{i}{\slsh p-m_i} =
  \delta^{ab}i\frac{\slsh p+m_i}{p^2-m_i^2}, \\
  \Delta_{g,\mu\nu}^{AB} &=& \delta^{AB}i\frac{-g_{\mu\nu}}{p^2}, \\
  \Gamma_{gq\bar{q}}^\mu &=& -i\,g_s\,t^A\,\gamma^\mu, \\
  \Gamma_{ggg} &=& -g_sf^{ABC}
  \left[(p-q)^\lambda g^{\mu\nu} + (q-r)^\mu g^{\nu\lambda}
    + (r-p)^\nu g^{\lambda\mu}\right].
\eee
Note that, apart from the triple-gluon vertex, the only difference
relative to QED is in the colour structure: propagators are diagonal in
colour and the vertex for a gluon of colour $A$ to scatter a quark of
colour $b$ to a quark of colour $c$ contains $(t^A)_{cb}$.  Note also
that unlike QED the quark--gluon vertex is flavour-independent (it is
straightforward to check that, unlike in QED, we cannot introduce a
flavour-dependence into the gauge transformation,
Eq.~(\ref{eq:localSUN}) and retain gauge invariance).  In the
triple-gluon vertex, the three gluons have momenta $p,q,r$, Lorentz
indices $\mu,\nu,\lambda$ and colour indices $A,B,C$ respectively.  The
momenta are all ingoing: $p+q+r=0$.

The Feynman rules for ghosts and for the four-gluon vertex can be found
in ESW\cite{ESW} (p.~10).  They will not be needed for this course.

Note also that in analogy with QED the strong charge $g_s$ is usually
substituted by~$\as$,
\be
  \as \equiv \frac{g_s^2}{4\pi}.
\ee

\subsection[$e^+e^-\to q\bar{q}$]{\boldmath$e^+e^-\to q\bar{q}$}
\label{sec:eeqq}

One of the most fundamental quantities in QCD is the total $e^+e^-$
annihilation cross section to hadrons.  We will see in a later lecture
that to leading order in $\as$ this is equal to the total $e^+e^-\to
q\bar{q}$ cross section.  The calculation is very similar to that for
$e^+e^-\to\mu^+\mu^-$, the only difference being in the colour
structure.  The photon is colour blind, so the Feynman rule for a photon
to couple to a quark contains a trivial colour matrix,~$\delta^{ab}$.
Summing over colours and dividing by the number of incoming colour
states (1 in this case since electrons are not coloured), we therefore
obtain
\be
  \sigma(e^+e^-\to q\bar{q}) = \sigma(e^+e^-\to\mu^+\mu^-)
  \times e_q^2 \times \sum_{a,b} \delta^{ab} \delta^{ba}.
\ee
We obtain
\be
  \sum_{a,b} \delta^{ab} \delta^{ba} = \sum_a \delta^{aa} = N_c,
\ee
and hence
\be
  R_{\mathrm{had}} \equiv
  \frac{\sigma(e^+e^-\to\mbox{hadrons})}{\sigma(e^+e^-\to\mu^+\mu^-)}
  = \sum_q e_q^2N_c.
\ee

\subsection[$e^+e^-\to q\bar{q}g$]{\boldmath$e^+e^-\to q\bar{q}g$}

This process will be important for the higher order corrections to
$\sigma(e^+e^-\to\mbox{hadrons})$ and particularly for the study of
three-jet final states in $e^+e^-$ annihilation, among the most
important test-beds for QCD.

\begin{figure}[t]
  \centerline{\hfill
    \includegraphics{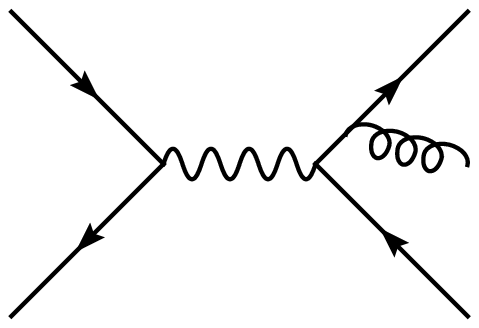}\hfill\hfill
    \includegraphics{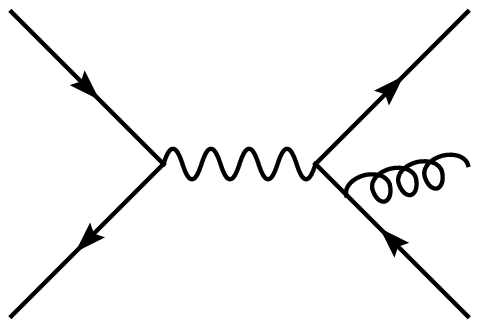}\hfill}
\caption[]{Feynman diagrams for the process $e^+e^-\to q\bar qg$}
  \label{fig:qqg}
\end{figure}

There are two Feynman diagrams, shown in Fig.~\ref{fig:qqg}.
We label the momenta and colours
$e^-(p_-)+e^+(p_+)\to q_a(p_1)+\bar{q}_b(p_2)+g_A(p_3)$.  For the matrix
element we obtain
\bee
  \hspace*{-0.5cm}
  i{\cal M} &=& \bar{v}(p_+)(ie)\gamma^\mu u(p_-)
  \; i\frac{-g_{\mu\nu}}s
  \; \varepsilon^{*\lambda}_A
\\&&\nonumber
  \bar{u}_a(p_1)\left\{
  (-ig_s)t_{ab}^A\gamma^\lambda
  \frac{\slsh p_1+\slsh p_3}{(p_1+p_3)^2}
  (-iee_q)\gamma^\nu +
  (-iee_q)\gamma^\nu
  \frac{-\slsh p_2-\slsh p_3}{(p_2+p_3)^2}
  (-ig_s)t_{ab}^A\gamma^\lambda
  \right\}v_b(p_2).
  \hspace*{-0.5cm}
\eee
We will evaluate the cross section from this matrix element later.  Here
we are interested in the colour algebra.  Using the fact that the spin
sum of a massless vector particle is proportional to the colour identity
matrix,
\be
  \sum_{\mathrm{spin}} \varepsilon^{*\mu}_A\varepsilon^\nu_B
  = -g^{\mu\nu}\delta_{AB},
\ee
we obtain
\be
  \sum|{\cal M}|^2 \propto
  \sum_{a,b,A} t^A_{ab} \left(t^A_{ab}\right)^* =
  \sum_{a,b,A} t^A_{ab} t^A_{ba} = \sum_A \Tr(t^At^A) =
  C_F \Tr(1) = C_FN_c,
\ee
where the first step uses the fact that $t^A$ are hermitian, the second
is simply a trivial rewrite, switching to matrix notation, the third
uses Eq.~(\ref{eq:colour2}) and the fourth uses the fact that the matrix
being traced is the identity matrix of the fundamental representation,
i.e.~the $N_c\times N_c$ identity matrix.  Note that since the colour
factor of the lowest order process is $N_c$, we can associate $C_F$ with
the emission of the additional gluon.  Since the emission probability of
a gluon from a quark is proportional to $C_F$, and we will later see
that that from a gluon is proportional to $C_A$, $C_F$ and $C_A$ are
sometimes referred to as the squares of the colour charges of the quark
and gluon respectively.

Performing the trace Dirac algebra on the matrix element, we finally
obtain
\be
  \label{qqgfinal}
  \sum|{\cal M}|^2 = \frac{16C_FN_ce^4e_q^2g_s^2}
  {s\,p_1\ldot p_3\,p_2\ldot p_3}
  \left((p_1\ldot p_+)^2+(p_2\ldot p_+)^2
       +(p_1\ldot p_-)^2+(p_2\ldot p_-)^2\right).
\ee
(Note the misprint in ESW~\cite{ESW} --- their result is a factor of 4 too
large.)

\subsection[The coupling constant $\as$ and renormalization]
 {\boldmath The coupling constant $\as$ and renormalization}
\label{renormalization}

As we mentioned above, in practical calculations, $\as$ is usually used
rather than $g_s$.  Besides the quark masses, which we will neglect in
most of this course, $g_s$ is the only parameter in the QCD Lagrangian
and therefore assumes a central role in our study of QCD.  However, it
is not {\it a priori\/} clear that parameters in the Lagrangian are
physically observable quantities --- any physical observable can be
calculated as a function of them (at least in perturbation theory) and
their values can be extracted from measured values of physical
observables, but they are not necessarily themselves physical.  It is
worthwhile therefore to consider whether we can reformulate our theory
in such a way that one physical observable can be written as a function
of another.  This reformulation is known as renormalization.

In this section I give a very handwaving description of renormalization,
which I believe conveys the important physical point.  Of course for
practical calculations one needs a much more precise definition of the
renormalization prescription, which I describe at the end.

\begin{figure}[t]
  \centerline{\hfill
    \includegraphics{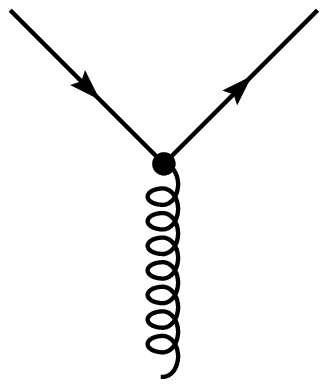}\hfill\hfill
    \includegraphics{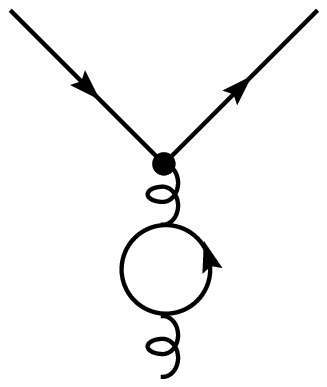}\hfill\hfill
    \includegraphics[106,579][195,686]{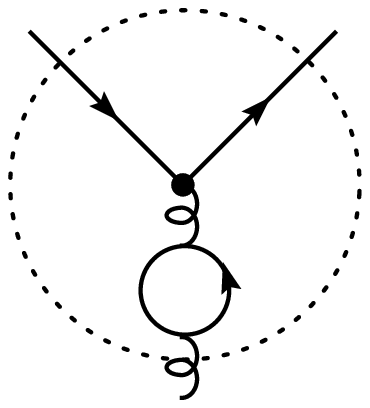}\hfill\hfill
    \includegraphics{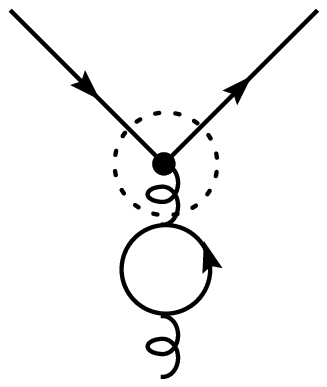}\hfill}
  \centerline{\hfill(a)\hfill\hfill(b)\hfill\hfill(c)\hfill\hfill(d)\hfill}
  \caption{When a quark--gluon vertex~(a) is corrected by a loop~(b),
  one must decide whether to describe it as a correction to the
  vertex~(c), or to the rest of the diagram~(d)}
  \label{fig:renorm}
\end{figure}

We redefine $g_s$ to be the strength of the quark--gluon coupling, as in
Fig.~\ref{fig:renorm}a.  At first sight, this seems like a trivial
statement and at the lowest order of perturbation theory it is~--- the
two definitions are identical.  However, when we calculate higher orders
of perturbation theory, we encounter loop corrections like the one in
Fig.~\ref{fig:renorm}b, which correct the vertex.  To avoid
double-counting, we must uniquely decide whether these corrections are
part of the vertex, as in Fig.~\ref{fig:renorm}c, or the rest of the
diagram, as in Fig.~\ref{fig:renorm}d.
One way to decide is to introduce a {\it renormalization scale\/}
$\mu_R$ and say that physics at high scales (therefore short distances)
above $\mu_R$ is part of the vertex and physics at lower scales (longer
distances) below $\mu_R$ is part of the rest of the diagram.  Of course,
this is simply a book-keeping device, which does not change the
physics, it simply ensures that each physical contribution to the
process is counted once and only once.  Since $\mu_R$ is a completely
arbitrary book-keeping scale, introduced by hand, its value should not
affect the physical prediction~--- changing it simply moves contributions
between what we call the vertex and what we call the rest of the
diagram.  Since the amount of physics that we include in the vertex
depends on $\mu_R$, and we defined $g_s$ to be the strength of the
vertex, it is clear that $g_s$ must now be a function of $\mu_R$.

It is worth mentioning that, although I defined $g_s$ as the strength of
the quark--gluon vertex, I could equally well have defined it as the
strength of the triple-gluon vertex.  It is one of the remarkable
features of gauge theories that, as a direct result of the gauge
symmetry, I would get exactly the same result for the renormalized
coupling $g_s(\mu_R)$.  That is, the equality of the strengths of the
quark--gluon and triple-gluon vertices is true even after
renormalization.

When it is clear that I am talking about the renormalization scale, I
will henceforth drop the subscript ${}_R$.

\subsubsection{Renormalization group equation}

As I said, varying $\mu$ moves physical contributions (loop corrections)
around within a calculation, but it should not change the result of the
physical calculation.  We can use this fact to derive an equation for
how $g_s$ varies as a function of $\mu$.  This is one of a set of
equations that together describe how the whole theory varies with
renormalization scale (and scheme), which formally form a group.

We study this by considering a dimensionless physical observable $R$
that is a function of only one physical scale $Q^2$ (think of
$R_{\mathrm{had}}$ at $\surd s=Q$ for example).  Assume that this
observable is not sensitive to quark masses (we will return to this
point shortly).  After renormalization, $R$ can only be a function of
$Q^2$, $\mu^2$ and $\as(\mu^2)$.  By dimensional analysis, the only way
the dimensionless function $R$ can depend on the dimensionful variables
$Q^2$ and $\mu^2$ is through their ratio.  We can therefore write
\be
  R = R(Q^2/\mu^2,\as(\mu^2)).
\ee
We can use the fact that $R$, as a physical quantity, must be
independent of the value of~$\mu$, and the chain rule for partial
derivatives, to write
\bee
  \mu^2\frac{d}{d\mu^2}R(Q^2/\mu^2,\as)=0
  &=&\left[\mu^2\frac\partial{\partial\mu^2}
    +\mu^2\frac{\partial\as}{\partial\mu^2}\,\frac\partial{\partial\as}
    \right]R
  \\&\equiv&\left[\mu^2\frac\partial{\partial\mu^2}
    +\beta(\as)\frac\partial{\partial\as}
    \right]R \ , 
  \label{toberearranged}
\eee
i.e., $\beta(\as)\equiv\mu^2\frac{\partial\as}{\partial\mu^2}$.  There
are several points to note about this.
\begin{itemize}
\item A physical solution is provided by $R(1,\as(Q))$, i.e., by setting
  the renormalization scale equal to the physical scale in the problem.
\item $Q$-dependence of the physical quantity $R$ comes about only
  because of the renormalization of the theory and would not be present
  in the classical theory.  Thus measuring the $Q$-dependence of $R$
  directly probes the quantum structure of the theory.
\item By rearranging Eq.~(\ref{toberearranged}), one can derive the
  $\mu^2$ dependence of $\as$ from a calculation of $R$,
\be
  \beta(\as) = -\frac{\mu^2\frac{\partial R}{\partial\mu^2}}
       {\frac{\partial R}{\partial\as}}.
\ee
\item If $\as$ is small, $R$ is perturbatively calculable and hence
  $\beta(\as)$ is too.
\end{itemize}
The $\beta$ function of QCD is now known to four-loop accuracy,
\be
  \beta(\as)=-\as^2(\beta_0+\beta_1\as+\beta_2\as^2+\beta_3\as^2+\ldots).
\ee
Although the higher orders are essential for quantitative calculation,
they are not for qualitative understanding: almost all QCD phenomenology
can be understood using the one loop result,
\be
  \beta_0 = \frac{11C_A-4T_RN_f}{12\pi},
\ee
where $N_f$ is the number of quark flavours that can appear in loops, to
be discussed further shortly.

Note that $\beta_0$ is positive and hence that the $\beta$ function is
negative, at least when $\as$ is small.  This results in asymptotic
freedom: the fact that the interactions become weak at high energies
(short distances) and infrared slavery: the fact that they become strong
at low energy.

If we neglect the higher orders, we can solve the renormalization group
equation exactly, to obtain $\as$ at some scale $Q$ as a function of its
value at the renormalization scale $\mu$,
\be
  \alpha_s(Q^2) = \frac{\alpha_s(\mu^2)}{1+\alpha_s(\mu^2)
    \beta_0\ln\frac{Q^2}{\mu^2}}.
  \label{asmu}
\ee

\subsubsection{Choosing $\mu^2$}

Although physical quantities do not depend on $\mu$, a calculation
truncated at a finite order of perturbation theory does.  We must
therefore choose some value for $\mu$.  To illustrate this, suppose that
our dimensionless physical quantity $R$ has a perturbative expansion
that starts at $\mathcal{O}(\as)$,
\be
  R=R_1\as+\ldots,
\ee
then if we truncate at leading order,
\be
  R\approx R_1\as,
\ee
our truncated expression for $R(1,\as(Q))$ can be expanded as a power
series in $\as(\mu^2)$
\bee
R(1,\as(Q^2))&\approx& R_1\as(Q^2)\\
&=&R_1\as(\mu^2)\left[
1-\beta_0\as(\mu^2)\ln\frac{Q^2}{\mu^2}+
\beta_0^2\as^2(\mu^2)\ln^2\frac{Q^2}{\mu^2}
+\ldots
\right].
\eee
The leading order result in renormalized perturbation theory is the
first term of this series, i.e., $R_1\as(\mu^2)$.  It is therefore clear
that although $\mu$ is completely arbitrary, choosing it far from $Q$
guarantees a large truncation error (note that the converse is not
true).  One should therefore choose $\mu^2$ `close' to $Q^2$, but how
close is close?

The conventional approach is to set $\mu=Q$ and to use the $\mu$
variation in a reasonable range, e.g., $Q/2$ to $2Q$ as an estimate of
the truncation uncertainty.  It should be clear from the foregoing
discussion that this is an extremely arbitrary procedure.  However, the
folklore is that in almost all cases where higher order corrections have
been calculated, they have fallen within the band given by this
procedure.

\subsubsection{Measuring $\as$}
\label{sec:measureas}

\begin{figure}[t]
  \centerline{\rotatebox{90}{\scalebox{0.5}{\includegraphics[118,106][525,678]{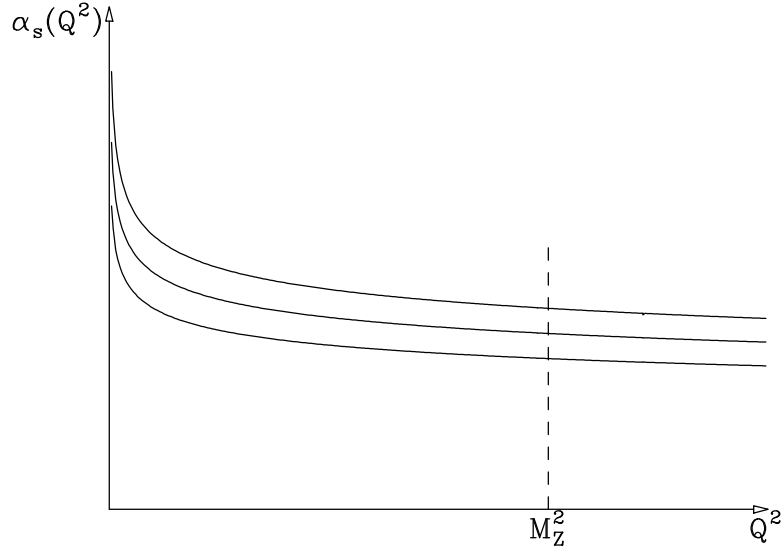}}}}
  \caption{A measurement of $\as$ at any scale $Q$ fixes which curve our
  universe lies on, but to compare measurements at different scales we
  have to agree to label the curves in a standard way, for example using
  $\as(M_z)$}
  \label{fig:alphas}
\end{figure}

The $\beta$ function tells us how $\as$ varies with scale, but it does
not tell us the value of $\as$ at any particular scale: we need an
experimental measurement to do that.  Effectively $\beta(\as)$ defines a
family of curves, as illustrated in Fig.~\ref{fig:alphas}, and one
measurement at any scale is sufficient to tell us which curve our
universe lies on.
However, in order to compare and combine measurements of $\as$ at
different scales, we have to agree on some convenient labeling of the
curves.  The measurement at any given scale can then be converted into a
measurement of the label.  Historically, this was often done using the
`QCD scale', $\Lambda_{QCD}$, described in the next section, but more
recently it has been realized that the value of $\as$ at some fixed
scale at which it is relatively small is a lot more convenient.  Since
some of the best measurements come from $Z^0$ decays, it has become
universal to use $\as(M_z)$ as the label.  We will discuss the
measurements of $\as$ further in Section~\ref{sec:asmeasure}.

\subsubsection{The `QCD Scale', $\Lambda$}

As I just mentioned, this is another way to label the running coupling,
which is to construct a renormalization group invariant scale from
$\as(\mu)$.  Although the Lagrangian of massless QCD has no scale, the
renormalization process introduces a dimensionful parameter,
\be
  \label{lambda_def}
  \Lambda^2 = \mu^2\exp-\int^{\as(\mu^2)}\frac{dx}{\beta(x)}
  \approx \mu^2\mathrm{e}^{-1/\beta_0\as(\mu^2)},
\ee
where the approximation uses only the one-loop term in the $\beta$
function%
\footnote{Note that the definition in the first equality of
  Eq.~(\ref{lambda_def}), while formally renormalization group
  invariant, is not practically useful, since the lower limit of the
  integration is not defined (corresponding to the fact that any
  definition of $\Lambda$ that differs by a multiplicative constant is
  equally renormalization group invariant).  For perturbative
  calculations, various definitions, equivalent to
  Eq.~(\ref{lambda_def}) to the order to which they are defined, can be
  used.  For non-perturbative calculations, for example in lattice QCD,
  the precise definition is more critical.  A commonly-used convention
  (see for example \cite{Gockeler:2005rv}) is
  $\Lambda^2 = \mu^2\exp\left\{-\frac1{\beta_0\as(\mu^2)}
  -\frac{\beta_1}{\beta_0^2}\log\as(\mu^2)
  -\int_0^{\as(\mu^2)}dx\left(\frac1{\beta(x)}+
  \frac{1-\frac{\beta_1}{\beta_0}x}{\beta_0x^2}\right)\right\}.$  In
  contrast to the definition given in~\cite{ESW}, for example, this can
  be seen to depend only on the $\beta$ function at $\as(\mu^2)$ and at
  smaller values, so is well-defined perturbatively and, as can be
  easily checked, is exactly renormalization scheme invariant.}%
.  This process by which a scaleless theory gets a physically
observable scale by the introduction of the unphysical renormalization
scale is known as dimensional transmutation.

At leading order, $\Lambda$ has a simple interpretation, it is the scale
at which the coupling becomes infinite.  However, this interpretation is
not self-consistent, since it relies on a truncation of the perturbation
series in a region in which the coupling is large, ultimately
divergent.  More generally, $\Lambda$ can be viewed as a renormalization
group invariant parameterization of the scale at which the theory
becomes non-perturbative.  All non-perturbative quantities, for example
the hadron masses, would be expected to be of order $\Lambda$.

However, $\Lambda$ is not a very practically useful label for the value
of $\as$.  This is because its precise value, for a given measured value
of $\as$, depends strongly on the theoretical input used in the
calculation, for example which order of perturbation theory we truncate
$\beta$ at, which renormalization scheme we use, the number of flavours
we assume, the way we match the running coupling at the flavour
thresholds, etc.

In principle any labeling suffers from these problems, but by using the
value of $\as$ in a region where it is small, and where the scale is not
too different from that at which the measurements are made, the impact on
$\as(M_z)$ is small, whereas $\Lambda$ is related to the region where
$\as$ is large, far away from where the measurements are made, and these
effects are large.

\subsubsection{Renormalization in practice}

To give a simple physical picture of renormalization, I have described
it in terms of a cutoff on the scale of the physical effects that are
included in different components of a Feynman diagram calculation.
However, in practice, this definition is extremely unattractive, because
it breaks Lorentz and gauge invariance, two of the fundamental
symmetries of our theory.  If calculating in this scheme, these
symmetries will get violated by a truncation at any finite order of
perturbation theory and only restored in an all orders calculation.
There are other simple schemes that work well in certain cases, for
example the Pauli--Villars regularization, but the only known scheme
consistent with all the symmetries of QCD, and hence guaranteed to work
at any order of perturbation theory, is {\it dimensional
regularization}.  In this section I give a very brief description of how
this works in practice.  The difference between $\mu$ and $\mu_R$ will
be (slightly) relevant here, so I temporarily reinstate the subscript.

The basic observation is that the loop corrections that we have been
discussing are divergent in four or more space-time dimensions, but are
finite in less than four dimensions.  We therefore choose to calculate
our Feynman diagrams in $d<4$ dimensions (we always work in Minkowski
space, with one time dimension and $d\!-\!1$ space dimensions).  With a
little thought, we can analytically continue the number of dimensions to
be a complex number such that at the end of the calculation, after the
renormalization prescription has been followed, we can let it smoothly
tend back to 4 and obtain finite results.  We therefore define
$d=4-2\epsilon$ and consider the $\epsilon\to0_+$ limit.

By counting the dimensionality of terms in the Lagrangian, we discover
that the coupling constant becomes dimensionful in $d\not=4$
dimensions.  This is not very convenient, so we define a dimensionless
parameter $\as$, by introducing a completely arbitrary scale~$\mu$,
\be
  \as^{(d)} = \as \; \mu^{2\epsilon},
\ee
where $\as^{(d)}$ is the dimensionful $d$-dimensional coupling.  $\mu$
is called the regularization scale.  It is often set equal to the
renormalization scale $\mu_R$, but I consider this confusing since we
have not yet renormalized the theory, so, for now, I keep them distinct
and only set them equal again at the end of this section.

When calculating loop corrections, we then find terms that have
$1/\epsilon$ singularities for small $\epsilon$.  These have the right
form to be absorbed by a redefinition (i.e.~a renormalization) of the
coupling.  Since we also want the renormalized coupling to be
dimensionless, we have to introduce a dimensionful scale at which the
renormalization is performed,~$\mu_R$.  To make this concrete, at
one-loop order, the prescription is straightforward: after calculating
all the one-loop diagrams, rewrite all occurrences of $\as$ in terms of
the renormalized coupling,
\be
  \as(\mu_R)=\as+\beta_0\,F(\epsilon)\,
  \left(\frac{\mu^2}{\mu_R^2}\right)^\epsilon\,\frac1{\epsilon}\,\as^2\,.
\ee
Provided $F(0)=1$, once this substitution has been made, the amplitude
is finite.  That is, the $\epsilon$ poles that this expression produces
exactly cancel those from the one-loop calculation.  Moreover, the
arbitrary scale $\mu$ cancels from the amplitude at this point.  One is
left with a finite amplitude that depends only on $\mu_R$ and
$\as(\mu_R)$, in exactly the same way as discussed earlier.

The arbitrary function $F(\epsilon)=1+\mathcal{O}(\epsilon)$ defines the
renormalization scheme.  More precisely, it defines what finite parts of
the loop amplitude are subtracted into the renormalized coupling, in
addition to the divergent part.  The MS, or minimal subtraction, scheme,
is defined by subtracting nothing else,
\be
  F_{\mathrm{MS}}(\epsilon)=1.
\ee
The most commonly used scheme is the $\mathrm{\overline{MS}}$, or
modified minimal subtraction, scheme, in which one identifies some
additional overall factors coming from the analytical continuation of
the angular integrations in the one-loop calculation.  Since they are
universal it is convenient to subtract them into the coupling,
\be
  F_{\mathrm{\overline{MS}}}(\epsilon)
  =\frac{(4\pi)^\epsilon}{\Gamma(1-\epsilon)}
  =1+(\ln4\pi-\gamma_E)\epsilon,
  \label{MSbar}
\ee
where $\Gamma$ is the Euler gamma function and $\gamma_E$ the Euler
gamma constant, $\gamma_E\approx0.577216$.  Note that the two
expressions on the right-hand side of Eq.~(\ref{MSbar}) differ at order
$\epsilon^2$.  Different practitioners use either of the two
definitions, resulting in a finite difference at two loops that is
straightforward to keep track of.

\subsection{Quark masses and decoupling}
\label{sec:decoupling}

The quark masses $m_q$ are also parameters of the Lagrangian and face
the same issues: for a physical calculation we should redefine them in a
physical way.  For the electron mass, we have a simple definition: we
can isolate a single electron and `weigh' it in the laboratory.  That is, we
can define its mass through the classical limit.  We cannot use the same
procedure for quarks, because confinement means that we can never take a
single quark off to our laboratory to weigh it individually.  We must therefore
define some other renormalization procedure.

It is possible to proceed in close analogy with the coupling strength.
We renormalize our theory at the same scale $\mu$.  We encounter gluon
loop corrections to the quark propagator and absorb the part of them at
scales above $\mu$ into the definition of the mass.  We therefore obtain
a `running' (i.e.~scale-dependent) mass.  Just like for the coupling, we
can obtain a renormalization group equation with
perturbatively-calculable coefficients,
\be
  \frac{\mu^2}{m}\,\frac{dm}{d\mu^2}=-\,\frac1\pi\,\as(\mu^2)+\ldots\,.
\ee
At leading order it can be solved exactly, to give
\be
  m(\mu^2) = M\left[\as(\mu^2)\right]^{\frac1{\pi\beta_0}},
\ee
where $M$ is a renormalization group invariant constant
(c.f.~$\Lambda_{QCD}$).  Note that increasing $\mu^2$ decreases~$m^2$.
Thus quarks appear to get lighter as they are probed at scales further
and further above their masses.

An alternative scheme, which is often used in electroweak physics, and
in the physics of heavy mesons, is the {\it pole mass}.  Here one
defines $m_q$ to be the pole of the propagator $i(\slsh
p+m_q)/(p^2-m_q^2)$ to all orders.  This is very useful for $Q\sim m_q$,
but it turns out that it is similar to a running mass scheme with $\mu$
of order $m_q$ and hence generates large logarithms and a large
truncation error for $Q\gg m_q$.

If our dimensionless observable $R$ is finite for massless quarks then
the quark mass effects must vanish smoothly as the mass goes to zero.
Therefore the mass effects must be suppressed by $(m_q/Q)^n$, with
$n\ge1$.  If there are quarks with mass much greater than $Q$, they can
only affect our observable through loop corrections.  A dimensional
argument shows that such corrections must be suppressed by $(Q/m_q)^n$,
with $n\ge2$.

These observations form the basis of the decoupling theorem, in which
quarks heavier than our physical scale can be ignored, and quarks
lighter than it can be treated as massless.  Thus, for most QCD
calculations, we work with $N_f$ flavours of massless quark (recall the
$N_f$ that appeared in $\beta_0$).  Care must be taken when $Q$ is close
to a quark mass, or we study a range of processes at scales that span a
quark mass, but in fact for most of the phenomenology considered in this
course we can simply take $N_f$ to be fixed, $N_f=5$.

\subsection{Summary}

We have seen that QCD is a gauge theory.  The fact that the gauge
symmetry is non-Abelian predicts that the gluon is self-interacting.
This leads to the fact that the theory becomes strongly interacting at
low energies, and hence non-perturbative, and weakly interacting at high
energies so that perturbation theory can be used.

The main tools that we will use to study QCD are the
{\it factorization\/} of non-pertur\-bative effects and the
{\it renormalization\/} and {\it decoupling\/} of high-energy physics.
These allow us to use perturbation theory and, in particular, the
Feynman rules, to study the phenomenology of QCD.

\section{QCD phenomenology at tree level}
\setcounter{equation}{0}
\setcounter{footnote}{0}
\setcounter{figure}{0}

Leading order perturbation theory, together with the one-loop
renormalization group equation is enough to understand a wide variety of
QCD phenomenology.  In this section, we briefly review the phenomenology
of QCD before introducing the complications of loop corrections to it in
the following section.  Most of the salient ideas are introduced in the
context of $e^+e^-$ annihilation and deep inelastic scattering, but
apply equally well to hadron collisions and photoproduction, which we
discuss more briefly at the end.

\subsection[The cross section for $e^+e^-\to$ hadrons]
           {The cross section for \boldmath$e^+e^-\to$ hadrons}

One of the most striking features of $e^+e^-$ annihilation events is the
fact that many of them produce many hadrons.  In trying to calculate the
cross section for this process, however, we are immediately faced with a
problem: the Lagrangian does not contain any information about hadrons,
so there are no Feynman rules involving them.  Even if there were,
calculating all the diagrams for events involving thirty or forty
particles would be prohibitively complicated, let alone integrating them
over the corresponding phase space to produce a total cross section.
Fortunately a simple application of the Feynman rules of QED, together
with some simple symmetry arguments, allows us to make a surprisingly
strong statement about the cross section for $e^+e^-$ annihilation to
hadrons.

We postulate that the matrix element for the sum of all diagrams in
which a virtual photon with Lorentz index $\nu$ and momentum $q$
produces a particular set of $n$ hadrons with momenta $\{p_1 \ldots
p_n\}$ is known and parameterize it by a function $T_\nu(n,q,\{p_1
\ldots p_n\})$.  Using this function, it is straightforward to write
down the matrix element for the full process,
\be
  {\cal M} = \left\{\bar{v}(q_2) e\gamma_\mu u(q_1)\right\}
\; \frac{-g^{\mu\nu}}{q^2} \; T_\nu(n,q,\{p_1 \ldots p_n\})
\ee
and hence the phase-space integral for its total cross section.  The
total cross section to produce any number of any type of hadrons is then
simply given by the sum of this integral over hadron type and
multiplicity (both generically represented by $\sum_n$),
\bee
\sigma &=& \frac{1}{2s} \; \frac14 \;
  \frac{e^2}{s^2}
\mathrm{Tr}(\slsh q_2\gamma^\mu \slsh q_1\gamma^\nu)
\\&&
\times\sum_n \int dPS_n\;
T_\mu(n,q,\{p_1 \ldots p_n\})\;
T^*_\nu(n,q,\{p_1 \ldots p_n\}).
\eee
We then define a new two-index tensor, $H_{\mu\nu}$, to represent this
sum of integrals,
\be
H_{\mu\nu}(q) \equiv \sum_n \int dPS_n \;
T_\mu \; T^*_\nu,
\ee
which after the integration and summation can only be a function of
$q$\footnote{Can you spot the flaw in this argument?  It assumes that
all information about the hadron momenta is washed out by the
integration, which is only true if they are massless.  In general since
$p_h^2$ is fixed at $m_h^2$ during the integration, $H$ also depends in
a complicated way on the masses of all possible hadrons.  In fact we
will shortly justify, on the basis of a space-time picture, neglecting
these, in the limit that $q^2$ is much greater than all $m_h^2$.  It
also ignores any other masses in the problem, like the Z mass, which we
remedy later on.}.  Now, there are only two possible Lorentz covariant
two-index tensor functions of one four-vector, $g_{\mu\nu}$ and $q_\mu
q_\nu$.  We therefore parameterize $H_{\mu\nu}$ as a linear combination
of these, with coefficients that are functions of the only available
Lorentz scalar, $q^2$,
\be
H_{\mu\nu} = A(q^2)g_{\mu\nu} + B(q^2)q_\mu q_\nu.
\ee
Finally, since the theory is gauge invariant (in practice boiling down
to invariance under the change $\varepsilon^\mu\to\varepsilon^\mu+q^\mu$
for the polarization vector of a photon of momentum $q$), $H_{\mu\nu}$
must be perpendicular to both $q^\mu$ and $q^\nu$,
\be
q^\mu H_{\mu\nu}=q^\nu H_{\mu\nu}=0,
\ee
giving a relation between the two functions,
\be
  A=-q^2B.
\ee
The final step is to realize that $B(s)$ has to be dimensionless.  Since
it is a function of only one dimensionful parameter, it must therefore
be constant.  We therefore have the fundamental prediction that (for
energies well above all hadron masses) the cross section to produce any
number of hadrons is proportional to that to produce a muon--antimuon
pair,
\be
R(e^+e^-) \equiv
\frac{\sigma(e^+e^-\to\mbox{hadrons})}{\sigma(e^+e^-\to\mu^+\mu^-)} =
\mbox{constant},
\ee
without knowing anything about the interactions of hadrons!

\begin{figure}[t]
  \centerline{\includegraphics*[266,287][482,452]{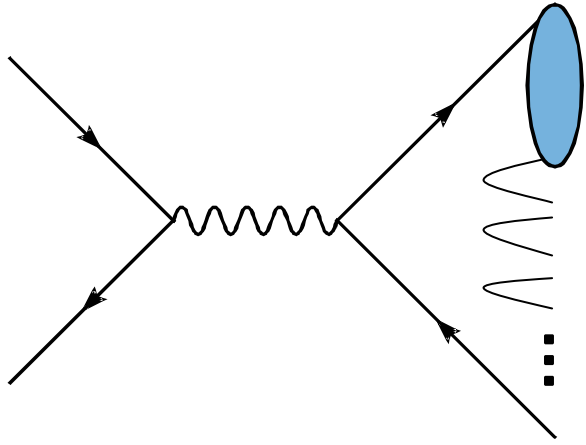}}
  \caption{Space-time sketch of the production of a hadron in $e^+e^-$
    annihilation}
  \label{fig:spacetime}
\end{figure}

In order to go further than this and try to predict this constant, or
learn something from its measurement, we need a specific model of the
production of hadrons.  This is provided by the {\it quark parton
  model}.  Of course this can be more rigorously derived, but I find it
more useful to illustrate the physics with a space-time argument, see
Fig.~\ref{fig:spacetime}.
Since the photon is highly virtual, it is produced and decays to quarks
in a small space-time volume, $t\sim1/\surd s$.  On the other hand, the
wavefunction of a hadron with mass $\sim m_{\mathrm{had}}$ has spatial
extent $\sim1/m_{\mathrm{had}}$ and hence the confinement of a quark
pair into the hadron takes $t\sim1/m_{\mathrm{had}}$.  Thus there is no
time for the confinement to affect the annihilation cross section and we
expect
\be
 \sigma(e^+e^-\to\mbox{hadrons}) \approx
 \sigma(e^+e^-\to\mbox{quarks}),
\ee
and the Feynman rules do tell us how to calculate that.

In fact, we can go further than that and use an argument from quantum
mechanics to postulate the form of the corrections to this
approximation.  Over a region of size $\sim1/\surd s$, the amount by
which the wave function of a hadron with spatial extent
$\sim1/m_{\mathrm{had}}$, could vary is $\sim m_{\mathrm{had}}/\surd s$
and the corrections should be at least this to some positive power,
\be
 \sigma(e^+e^-\to\mbox{hadrons}) =
 \sigma(e^+e^-\to\mbox{quarks})
\times\left(1+
{\cal O}\left(\frac{m_{\mathrm{had}}}{\sqrt{s}}\right)^n\right).
\ee
On the basis of the space-time picture, we can only justify that the
corrections to the quark parton model are suppressed by some (positive)
power of the ratio of scales.  In practice, $n$ is believed to be 6 for
$e^+e^-$ annihilation, making these corrections so small as to be almost
impossible to measure.  For most cross sections however, $n$ is~2, and
for jet cross sections,~1.

We calculated the cross section for $e^+e^-\to q\bar{q}$ in
Section~\ref{sec:eeqq} and obtained
\be
  R_{e^+e^-} \equiv \frac{\sigma(\mathrm{hadrons})}{\sigma(\mathrm{muons})}
  = N_c \sum_q e_q^2,
\ee
where the sum over $q$ is over all quark flavours that are kinematically
allowed, i.e.~for which $\surd s>2m_q$.  If we ignore effects close to
threshold, such as the formation of bound states, we can expect a plot
of $R_{e^+e^-}$ against $\surd s$ to present a series of steps at twice
the quark masses and be flat in between.  In principle one can read off
the quark masses and charges from this plot.

\begin{figure}[tp]
  \centerline{\scalebox{0.64}{\includegraphics{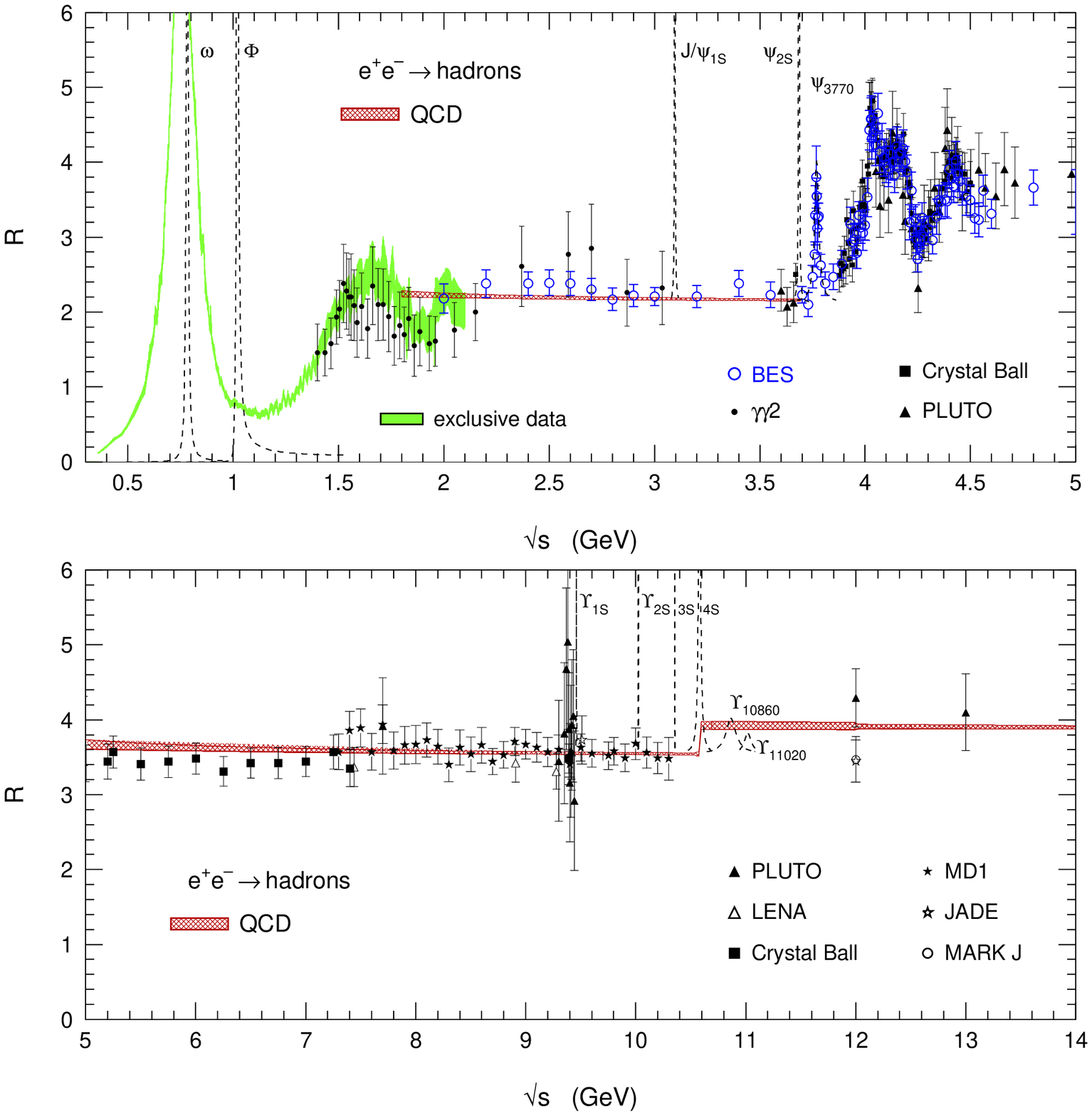}}}
  \centerline{\rotatebox{90}{\scalebox{0.7}{\includegraphics*[71,63][265,546]{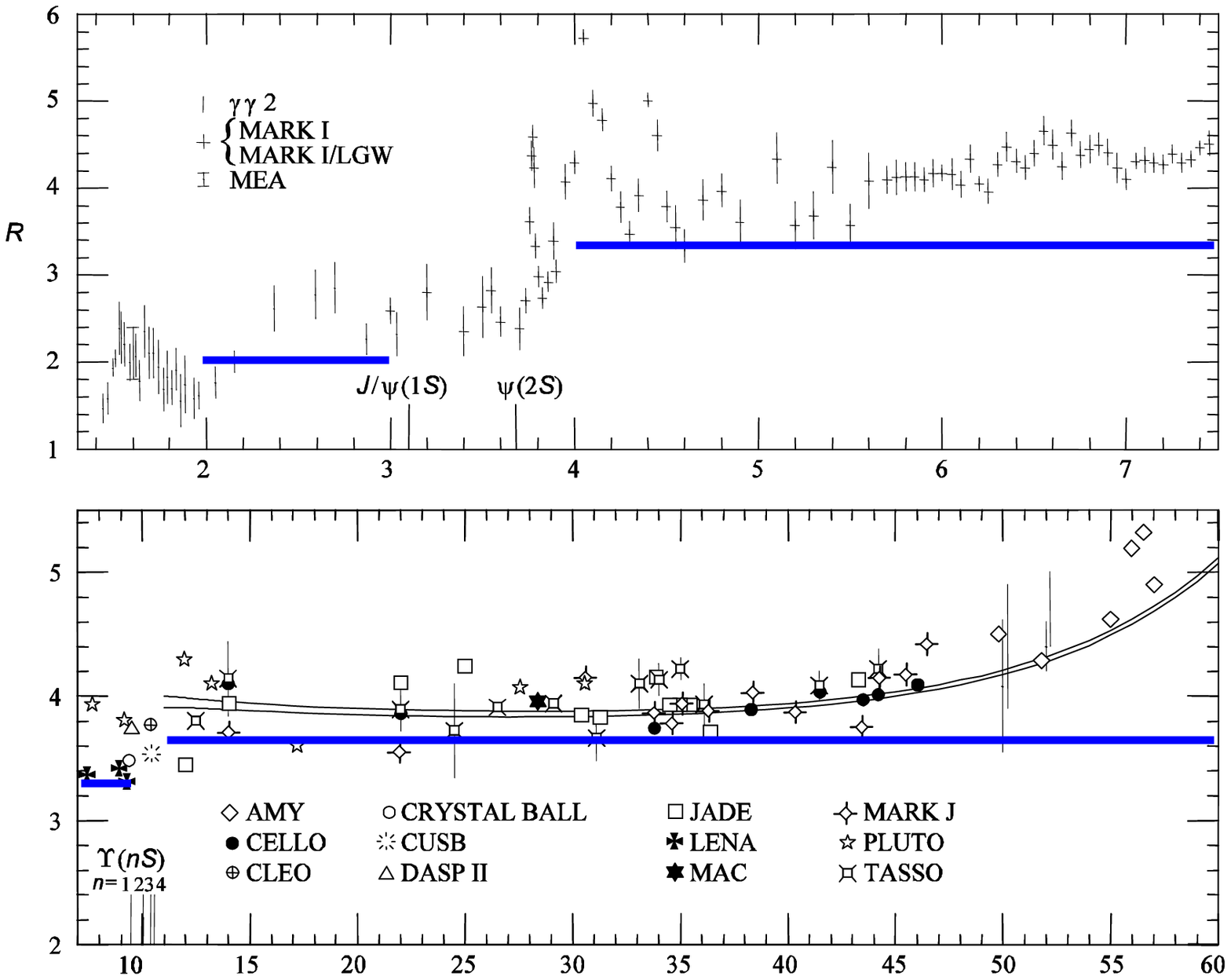}}}}
  \caption{Data on $R_{e^+e^-}$ as a function of centre-of-mass energy.
    Upper two panels taken from \cite{Davier:2002dy}, lower from
    ESW~\cite{ESW}.  The bands (red above, white below) show the QCD
    prediction, while the horizontal lines in the lower panel show the
    quark parton model expectations.}
  \label{fig:Re+e-data}
\end{figure}

Looking at the data in Fig.~\ref{fig:Re+e-data}, we see that the general
trend is as expected, but there are clearly corrections that are not
accounted for by the quark parton model.
One of these is the effect of higher order QCD corrections, which we
include in the next lecture.  Another is the effect of the $Z^0$ boson,
which is clearly seen at the high energy end of
Fig.~\ref{fig:Re+e-data}, which we include shortly.

Before including the $Z^0$ contribution, it is worth remarking on a
historical ambiguity that affects this figure.  Although people wrote
\[
R \equiv
\frac{\sigma(e^+e^-\to\mbox{hadrons})}{\sigma(e^+e^-\to\mu^+\mu^-)}
\]
they often didn't actually use that formula to show their experimental
results, but rather
\[
R \equiv \frac{\sigma(e^+e^-\to\mbox{hadrons})}
{\frac{4\pi\alpha^2}{3s}},
\]
using the leading order QED result for the denominator.  Clearly many of
the experimental and theoretical systematic errors would be smaller if
the former was used, although of course the statistical errors would be
larger, by around a factor of 2.  More recent measurements, for example
from LEP, have used the more honest notation in which the numerator and
denominator are calculated or measured in the same way.  This is
sometimes called $R_{had}$ to differentiate it from $R$.

\begin{figure}[t]
  \centerline{\rotatebox{90}{\scalebox{0.5}{\includegraphics{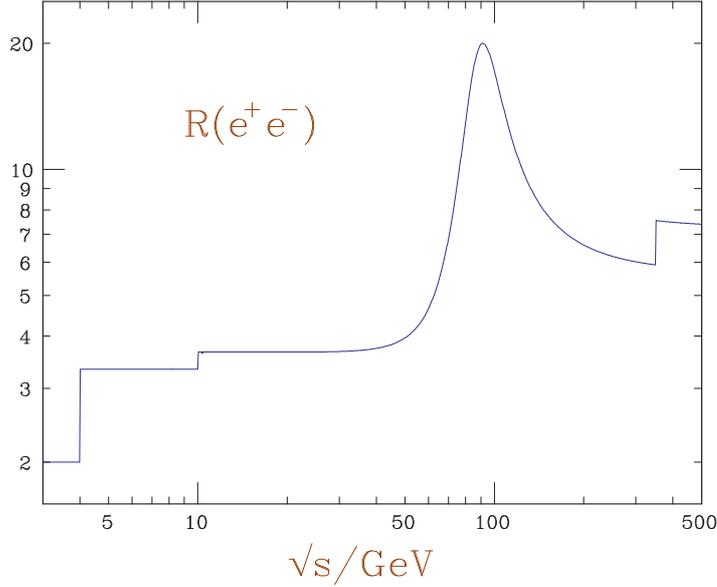}}}}
  \caption{Calculation of $R_{had}$ as a function of centre-of-mass
    energy}
  \label{fig:Re+e-calc}
\end{figure}

In Fig.~\ref{fig:Re+e-calc} I show the calculation of $R_{had}$ in the
quark parton model, including the $Z^0$ contribution.
It is clear that $\gamma$--Z interference is important, even far from
the Z peak.  However, exactly on the peak the interference is zero (you
might like to think about a simple explanation for why) and $R_{had}$ is
given to a good approximation by the Z contribution alone,
\be
  R_{had}=N_c\frac{\sum_q{v_q^2+a_q^2}}{v_\mu^2+a_\mu^2}=20.095,
\ee
where $v_i$ and $a_i$ are the vector and axial couplings of the $Z^0$ to
fermion type $i$.  I note for future reference that the value including
the photon contribution is 19.984.  This number compares well with the
LEP average measured value of $20.767\pm0.025$.  However, the difference
is still large on the scale of the experimental uncertainty, again
indicating a clear need for the QCD corrections.

\subsubsection{$\tau$ decays}

\begin{figure}[t]
  \centerline{\includegraphics{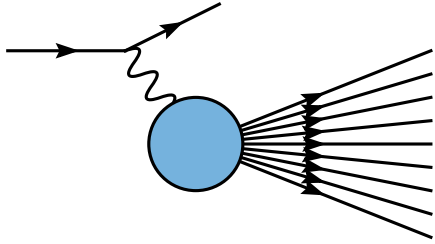}}
  \caption{Decay of $\tau$ lepton to hadrons}
  \label{fig:taudec}
\end{figure}

We conclude this section by mentioning the closely-related process of
$\tau$ decay to hadrons, depicted in Fig.~\ref{fig:taudec}.
One can apply exactly the same arguments to the blob in this diagram as
to annihilation of $e^+e^-$ to hadrons.  The only differences are that
we have a virtual W boson producing hadrons instead of a virtual photon,
and that we have an integral over all virtualities of the W between the
$\tau$ mass and zero, rather than a single virtuality fixed by the beam
energies.  Nevertheless exactly the same arguments follow through and
one obtains
\be
  R_\tau \equiv
  \frac{\mathcal{B}(\tau\to\mbox{hadrons})}{\mathcal{B}(\tau\to\mu)}
  =N_c\sum_{i,j}|V_{ij}|^2\approx N_c,
\ee
where the sum is over the flavours of quark and antiquark that can
appear in the W decay and $V$ is the CKM matrix.  Since a $\tau^-$ can
decay to $\bar ud$ or $\bar us$, to a good approximation this sum is
$\cos^2\theta_C+\sin^2\theta_C$ and the final result follows.

We will see later that this process provides an excellent measurement of
$\as$.



\subsection{Deep inelastic scattering}

Historically, the quark model developed as a way of rationalizing the
vast array of strongly-interacting particles that had been found by the
1960s.  However, it was not clear whether quarks were really physical
constituents of hadrons, or merely a convenient mathematical language to
describe the hadrons' wave functions.  The decisive evidence came from
deep inelastic scattering experiments at SLAC.  Today, deep inelastic
scattering experiments give us by far the best information about the
internal structure of the proton.

\subsubsection{Quarks as partons in hadronic scattering}

The classic probe of nuclear structure is electron--nucleus scattering.
Assuming the scattering takes place by exchanging a single photon,
measuring the kinematics of the scattered electron uniquely constrains
that of the photon.  The scattered electron has two non-trivial
kinematic variables, its energy and scattering angle.  These can more
conveniently be converted into the photon virtuality
($Q^2\equiv-q\ldot q$) and energy in the nucleus rest frame $\nu$.
$Q^2$ controls the resolving power of the photon, $Q^2\sim1/\lambda^2$.
For fixed small $Q^2\ll1/R^2$, where $R$ is the nuclear radius, the
photon is absorbed elastically by the nucleus, giving a narrow peak in
the $\nu$ distribution at $\nu=Q^2/2M_N$.  For increased $Q^2\sim1/R^2$
one begins to resolve nuclear resonances as additional peaks at higher
$\nu$.  Finally, for large $Q^2\gg1/R^2$, one resolves the proton
constituents of the nucleus, with the photon being absorbed elastically
by individual protons.  These show up as a peak at $\nu=Q^2/2M_p$,
broadened by the internal motion of the proton within the nucleus.

The scattering of electrons off hadrons, protons for example, is
exactly analogous: at low $Q^2$ one sees only elastic proton scattering,
but as $Q^2$ is increased, the photon can be elastically absorbed by the
(charged) quark constituents of the proton.  (Eventually at very large
$Q^2$ and $\nu$ something new happens relative to the nuclear case, but
we will not discuss that until the next lecture.)

We are interested in the region of Deep ($Q^2\gg M_p^2$) Inelastic
($W^2\gg M_p^2$, where $W$ is the invariant mass of the photon--proton
system) Scattering, DIS.  We are therefore justified in neglecting the
proton mass throughout, provided we do not work in the proton
rest-frame, which is not well defined in that case.  This is most
conveniently done by working in terms of Lorentz-invariant variables.

\subsubsection{Lorentz-invariant variables}

\begin{figure}[t]
  \centerline{\includegraphics{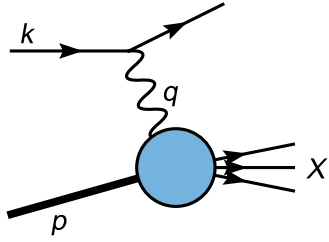}}
  \caption{Deep inelastic scattering}
  \label{fig:DIS2}
\end{figure}

\begin{figure}[t]
  \centerline{\scalebox{0.7}{\includegraphics{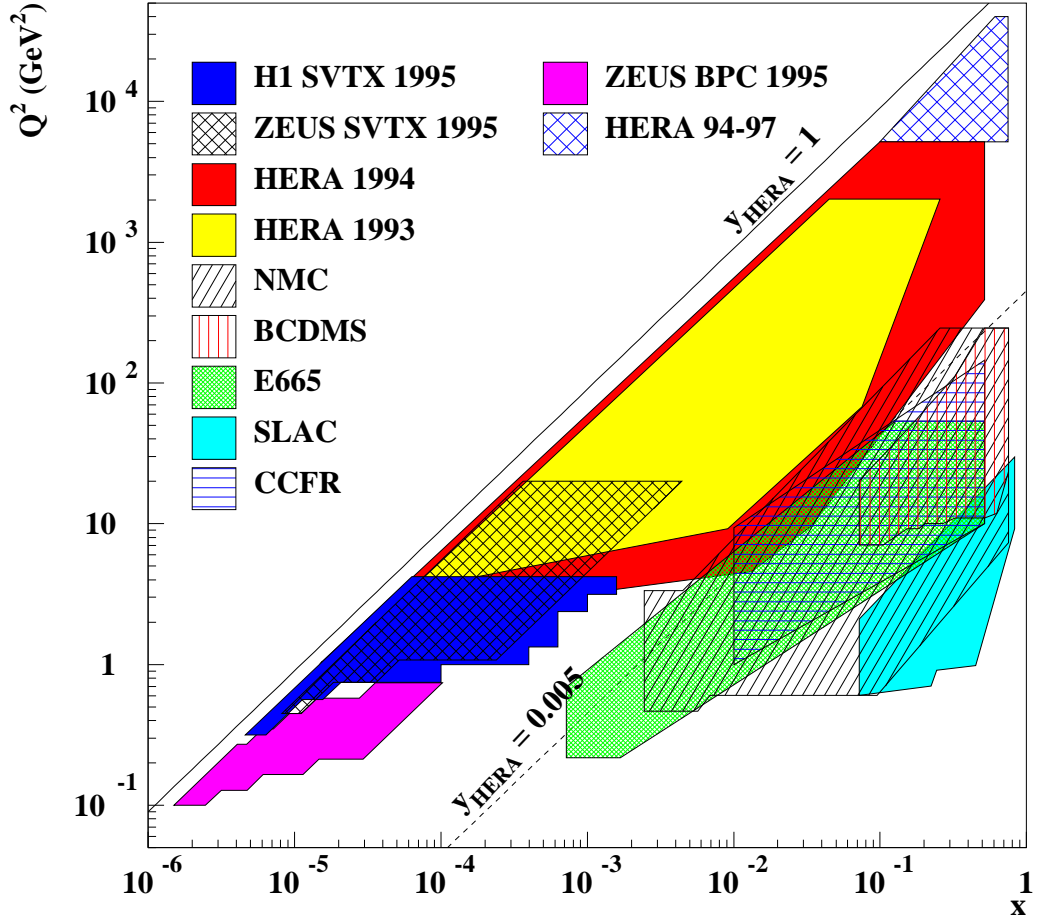}}}
  \caption{The $x$--$Q^2$ plane, showing the coverage of measurements by
  various experiments}
  \label{fig:hera}
\end{figure}

It is convenient to describe this in terms of Lorentz-invariant
variables.  We label the momenta as shown in Fig.~\ref{fig:DIS2}.
For an electron of momentum $k$ to scatter to one of momentum $k'$ by
exchanging a photon of momentum $q$ with a proton of momentum $p$ we
again have, for fixed centre-of-mass energy $s$, only two independent
kinematic variables,
\bee
  s &=& (k+p)^2, \\
  Q^2 &=& -q^2, \\
  x &=& \frac{Q^2}{2p\ldot q},
\eee
in terms of which we can calculate two other commonly-used variables
\bee
  W^2 &=& (p+q)^2 \;=\; Q^2\frac{1-x}x, \\
  y &=& \frac{p\ldot q}{p \ldot k} \;=\; \frac{Q^2}{xs}.
\eee
The kinematic limits are
\bee
  Q^2 < s, \\
  x > \frac{Q^2}s.
\eee
The coverage of the $(x,Q^2)$ plane by the HERA, and earlier fixed
target, DIS experiments is shown in Fig.~\ref{fig:hera}

\subsubsection{Structure functions}

Since we do not yet know anything about the internal structure of
protons, we cannot calculate the matrix element for the interaction of
a photon with the proton to produce some arbitrary state $X$.  However,
like in the case of $e^+e^-$ to hadrons we can get a surprisingly long
way just by considering the properties that that matrix element must
satisfy.

We parameterize the matrix element for a proton of momentum $p$ to
absorb a photon of momentum $q$ and Lorentz index $\mu$ to produce an
arbitrary set of hadrons $X$ with fixed momenta $\{p_X\}$ as
\be
  e \, T_\mu(p,q;\{p_X\}).
\ee
We therefore have the matrix element squared for the whole process
\be
  \frac14|{\cal M}|^2 = \frac14\,\frac{e^4}{Q^4}\,
  \Tr\left\{\slsh k \gamma^\mu \slsh k' \gamma^\nu \right\} \,
  T_\mu(p,q;\{p_X\}) \, T^*_\nu(p,q;\{p_X\}).
\ee
For convenience we define the Lorentz tensor
\be
  L^{\mu\nu} =
  \Tr\left\{\slsh k \gamma^\mu \slsh k' \gamma^\nu \right\}.
\ee

If the state $X$ consists of $n$ hadrons, then the $n\!+\!1$-body phase
space for the whole process can be factorized into a part describing the
electron kinematics times the $n$-body phase space for $X$,
\be
  \label{PSfact}
  dPS = \frac{Q^2}{16\pi^2 s x^2} \, dQ^2\,dx\,dPS_X.
\ee
This is as far as we can go for a specific state $X$, but we can get
further by integrating over the phase space of $X$ and summing over all
possible states $X$.  We define
\be
  \sum_X \int dPS_X \, \frac14|{\cal M}|^2 \equiv
  \frac{e^4}{Q^4} \, L^{\mu\nu}H_{\mu\nu},
\ee
or
\be
  \sum_X \int dPS_X \, T_\mu(p,q;\{p_X\}) \, T^*_\nu(p,q;\{p_X\}) =
  H_{\mu\nu}.
\ee
Since we have summed and integrated out all dependence on $X$,
$H_{\mu\nu}$ can only depend on the vectors $p$ and $q$.  Since the
electromagnetic and strong interactions conserve parity, it must be
symmetric in $\mu$ and $\nu$.  There are only four possible symmetric
two-index tensors that can be constructed from two vectors, so we can
parameterize the hadronic tensor as a linear combination of them:
\be
  H_{\mu\nu} = -H_1g_{\mu\nu} + H_2\frac{p_\mu p_\nu}{Q^2}
  + H_4\frac{q_\mu q_\nu}{Q^2} + H_5\frac{p_\mu q_\nu+q_\mu p_\nu}{Q^2},
\ee
where the $H$s are scalar functions of the only two Lorentz scalars
available $q\ldot q=-Q^2$ and $p\ldot q=Q^2/2x$, i.e., of $x$ and $Q^2$
only (not $s$).  (Note that we neglect $p\ldot p=M_p^2$ since we work
in the limit $|q\ldot q|,p\ldot q\gg p\ldot p$.)

If we include $Z^0$ exchange (or charged current scattering) we can
construct one further tensor, which is antisymmetric in $\mu$ and $\nu$,
$H_3\,\epsilon_{\mu\nu\lambda\sigma}p^\lambda q^\sigma$, where
$\epsilon_{\mu\nu\lambda\sigma}$ is the totally antisymmetric Lorentz
tensor.

Contracting with $L^{\mu\nu}$ we find that $H_4$ and $H_5$ cannot
contribute to physical cross sections (think about a simple explanation
why not) and we have
\be
  L^{\mu\nu}H_{\mu\nu} =
  4k\ldot k' \, H_1 + 4\frac{p\ldot k\;p\ldot k'}{Q^2} \, H_2.
\ee
Redefining (just a matter of convention) $H_1=4\pi F_1$ and
$H_2=8\pi xF_2$, we obtain the final result for the scattering cross
section
\be
  \frac{d^2\sigma}{dx\,dQ^2} = \frac{4\pi\alpha^2}{xQ^4}
  \left[y^2xF_1(x,Q^2) + (1-y)F_2(x,Q^2)\right].
\ee
Without knowing anything about the interactions of hadrons, we have been
able to derive the $s$ dependence of the scattering cross section for
fixed $x$ and $Q^2$ (which enters through the $y$ dependence: recall
$y=Q^2/xs$).

The $F$s are called the structure functions of the proton.  It is
common to see other linear combinations of the structure functions,
\bee
  F_T(x,Q^2) &=& 2xF_1(x,Q^2), \\
  F_L(x,Q^2) &=& F_2(x,Q^2) - 2xF_1(x,Q^2),
\eee
which correspond to scattering of transverse and longitudinally
polarized photons respectively.  We therefore have
\be
  \frac{d^2\sigma}{dx\,dQ^2} = \frac{2\pi\alpha^2}{xQ^4}
  \left[(1+(1-y)^2)F_T(x,Q^2) + 2(1-y)F_L(x,Q^2)\right].
\ee
In fact the most common form you will see this in nowadays is
\be
  \label{F2initial}
  \frac{d^2\sigma}{dx\,dQ^2} = \frac{2\pi\alpha^2}{xQ^4}
  \left[(1+(1-y)^2)F_2(x,Q^2) - y^2F_L(x,Q^2)\right].
\ee
For the majority of current data, $y^2$ is small and $F_L$ can be
neglected: only close to the kinematic limit, or for very precise data,
need it be considered.

We have isolated all the non-trivial $x$ and $Q^2$ dependence into the
two functions $F_2(x,Q^2)$ and $F_L(x,Q^2)$, but we still have no idea
how those functions behave.  If we make the assumption that the
interaction of the photon with the innards of the proton does not
involve any dimensionful scale, then we immediately get the result that
the dimensionless $F$s cannot depend on the dimensionful $Q^2$ and we
get
\be
  \frac{d^2\sigma}{dx\,dQ^2} = \frac{2\pi\alpha^2}{xQ^4}
  \left[(1+(1-y)^2)F_2(x) - y^2F_L(x)\right],
\ee
known as Bjorken scaling.  Experimentally this is true to a pretty good
approximation, but given that the proton is supposed to consist of
quarks, bound together with a distance scale $\sim1/M_p$, how can the
interaction possibly be $M_p$-independent?  The answer to this lies in
the parton model.

\subsubsection{Parton distribution functions and Bjorken scaling}

\begin{figure}[t]
  \centerline{\includegraphics*[145,325][588,434]{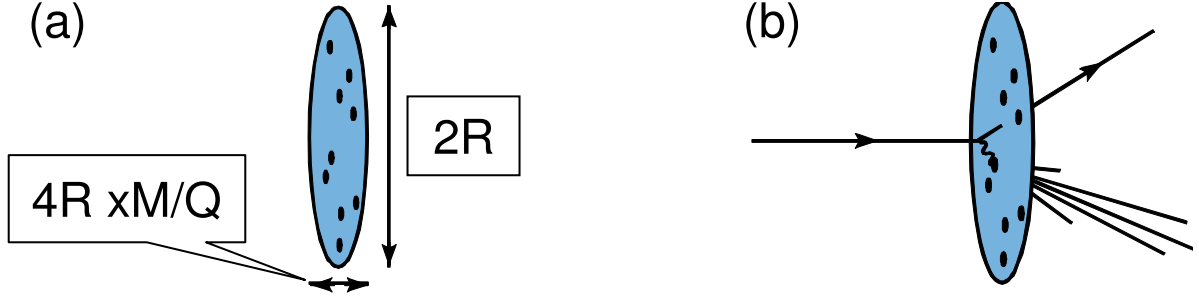}}
  \caption{In the Breit frame, the proton of diameter $2R$ is
  contracted to a pancake of thickness $4RxM_p/Q$ (a) so that a photon
  of high virtuality $Q$ interacts incoherently with a single parton
  within it (b)}
  \label{fig:Breit}
\end{figure}

Although it is of course Lorentz-invariant, the parton model is most
easily formulated in a frame in which the proton is fast moving.
Most convenient is the so-called Breit frame, in which the photon has
zero energy and collides head-on with the proton.  In this frame, the
proton energy is $Q/2x$.  Assuming that in its own restframe it is a
sphere of radius $R$, in the Breit frame it is massively Lorentz
contracted to a flat pancake, still with transverse diameter $2R$, but
with length $4RxM_p/Q\ll2R$, as illustrated in Fig.~\ref{fig:Breit}a.
The transverse size of the photon is $\sim1/Q\ll2R$.  The photon
therefore interacts with a tiny fraction of a thin disk, so provided
that the quarks are sufficiently dilute the photon is not able to
resolve the quarks' interactions and they act as if they were free.
That is, the photon effectively collides with a single free quark, as
illustrated in Fig.~\ref{fig:Breit}b.

Since they act as if they do not interact, their interactions do not
introduce a dimensionful scale, and so the structure functions will
obey Bjorken scaling.

More precisely, we suppose that the proton consists of a bundle of
comoving partons, which carry a range of the proton's momentum.  We
posit probability distribution functions (more often called parton
distribution functions, pdfs), such that partons of type $q$ carry a
fraction of the proton's momentum between $\eta$ and $\eta+d\eta$ a
fraction $f_q(\eta)d\eta$ of the time.  Provided that these partons are
pointlike $r^2\ll1/Q^2$ and dilute $f_q(\eta)\ll Q^2R^2$, the photons
will scatter incoherently off individual partons.  The cross section
can then be factorized as the convolution of the pdfs with the cross
section for parton scattering,
\be
  \frac{d^2\sigma(e+p(p))}{dx\,dQ^2} = \sum_q \int_0^1 d\eta \, f_q(\eta)
  \frac{d^2\sigma(e+q(\eta p))}{dx\,dQ^2}.
\ee
We will calculate the partonic cross section shortly, but first let me
point out a couple of features it must have.

Firstly if we assume that the scattering is elastic, then the outgoing
parton must be on mass-shell.  Since we are then considering a
two-to-two collision, which has only one nontrivial kinematic variable,
the double-differential cross section in $x$ and $Q^2$ must be
proportional to a $\delta$ function fixing one of the variables.
Specifically, if we assume that the partons are massless, then we obtain
the relation
\be
  (q+\eta p)^2 = 2\eta\,p\ldot q - Q^2 = 0,
\ee
or
\be
  \eta=x.
\ee

Secondly if we assume that the struck partons are the quarks of the
quark model, they must be fermions.  Simply from helicity conservation,
we can then show that $F_L=0$.  This is known as the Callan--Gross
relation and was one of the first proofs that the quarks of the quark
model really were the partons of the parton model.  (If the partons
were instead scalars we would have $F_T=0$ and hence completely
different $y$-dependence of the cross section.)

\subsubsection{Scattering cross sections}

To calculate the parton model prediction for the structure functions, we
need the matrix elements for $eq\to eq$.  These can be obtained by
crossing symmetry from those for $e^+e^-\to q\bar{q}$.  That is,
\be
  \sum|{\cal M}|^2 = 8(4\pi\alpha)^2\,e_q^2N_c\,
  \frac{(p_e\ldot p_q)^2+(p_e\ldot p_q')^2}{(p_e\ldot p_e')^2}.
\ee
Converting to the kinematic variables we defined earlier, we have
\be
  \sum|{\cal M}|^2 = 8(4\pi\alpha)^2\,e_q^2N_c\,
  \frac{1+(1-y)^2}{y^2}.
\ee
Using (\ref{PSfact}), we have
\be
  dPS = \frac{Q^2}{16\pi^2 s x^2}dQ^2 \, dx \, dPS_X.
\ee
Since $X$ consists only of one massless parton, we have
\bee
  dPS_X &=& \frac{d^4p_X}{(2\pi)^3}\delta(p_X^2)
  \,(2\pi)^4\delta^4(\eta p+q-p_X) \\
  &=& (2\pi)\delta((\eta p+q)^2) \\
  &=& \frac{2\pi x}{Q^2}\delta(\eta-x).
\eee
The full cross section is therefore
\bee
  \frac{d\sigma}{dx\,dQ^2} &=& \frac1{4N_c}\,\frac1{2\hat{s}}
  \,\frac{Q^2}{16\pi^2 s x^2}\,\frac{2\pi x}{Q^2}
  \,\delta(x-\eta)\,\sum|{\cal M}|^2 \\
  &=& \frac1{4N_c}\,\frac{y^2}{16\pi Q^4}
  \,\delta(x-\eta)\,\sum|{\cal M}|^2,
\eee
where the factor of $1/N_c$ is the average over incoming colours.  We
therefore have
\be
  \frac{d\sigma(e+q)}{dx\,dQ^2} = \frac{2\pi\alpha^2}{Q^4}
  \,\delta(x-\eta)\,\,e_q^2\,
  \left(1+(1-y)^2\right)
\ee
and hence
\be
  \label{F2final}
  \frac{d\sigma(e+p)}{dx\,dQ^2} = \frac{2\pi\alpha^2}{xQ^4}
  \left(1+(1-y)^2\right)\sum_q e_q^2\,xf_q(x).
\ee
Comparing (\ref{F2final}) with (\ref{F2initial}) we therefore have
\bee
  F_2(x,Q^2) &=& \sum_q e_q^2\,xf_q(x), \\
  F_L(x,Q^2) &=& 0.
\eee
Note that $F_2$ is $Q^2$-independent, showing Bjorken scaling.

\begin{figure}[t]
\vspace{.9cm}
  \centerline{\rotatebox{270}{\scalebox{0.5}{\includegraphics{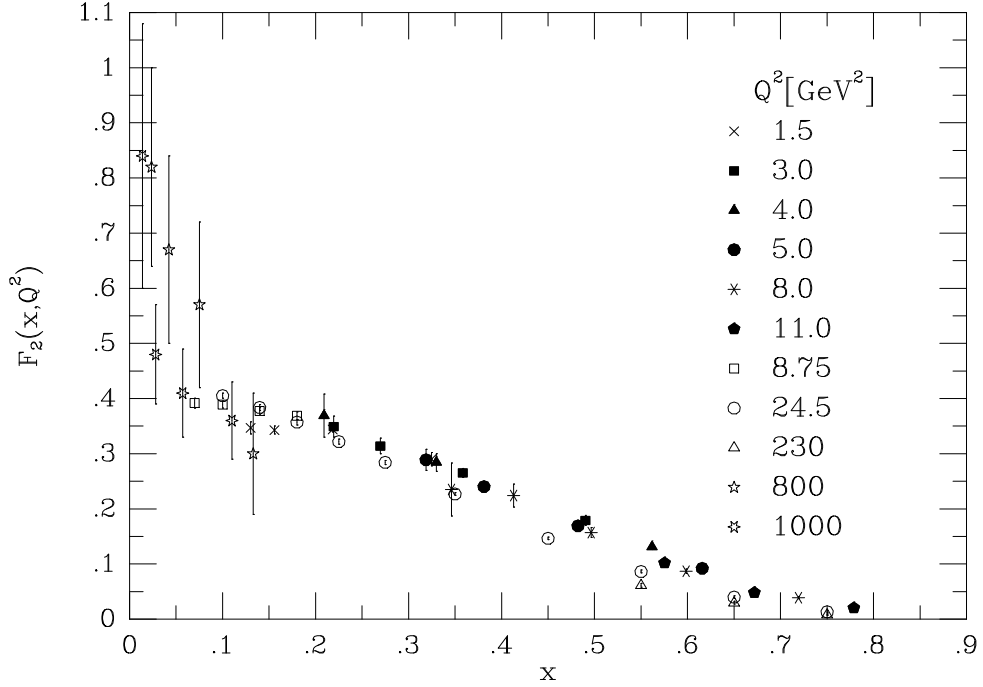}}}}
\vspace{.2cm}
  \caption{The structure function $F_2$ as a function of $x$ for various
  $Q^2$ values, exhibiting Bjorken scaling, taken from ESW~\cite{ESW}}
  \label{fig:bjscaling}
\end{figure}

Although we will see that QCD corrections do violate Bjorken scaling, it
is satisfied pretty well by the data, as can be seen in
Fig.~\ref{fig:bjscaling}.

\subsubsection{Charged current neutrino DIS}

We can consider charged current neutrino scattering in exactly the same
way.  Since the scattering takes place by the weak interaction, parity
is violated, allowing one additional Lorentz structure,
\bee
  L_{\mu\nu}^{\nu\atop\bar\nu} &=& L_{\mu\nu}^{e} \pm
  2i\epsilon_{\mu\nu\rho\sigma}k^\rho k'^\sigma, \\
  H^{\mu\nu} &=& -H_1g^{\mu\nu} + H_2\frac{p^\mu p^\nu}{Q^2} -
  \frac{i}{Q^2}\epsilon^{\mu\nu\rho\sigma}p_\rho q_\sigma H_3, \\
  \Rightarrow
  L_{\mu\nu}^{\nu\atop\bar\nu}H^{\mu\nu} &=&
  2Q^2H_1 + Q^2\frac{1-y}{x^2y^2}H_2
  \pm \frac{Q^2}{xy}H_3\;(1-y/2).
\eee
Thus, defining $H_3=8\pi xF_3$, we have a third structure function
$F_3$:
\be
  \frac{d^2\sigma({\nu\atop\bar\nu}+p)}{dx\,dQ^2}
  = \frac{G_F^2}{4\pi x}\left(\frac{M_w^2}{Q^2+M_w^2}\right)^2
    \left[\left(1+(1-y)^2\right)F_2^{\nu\atop\bar\nu}
    -y^2F_L^{\nu\atop\bar\nu}
    \pm\left(1-(1-y)^2\right)xF_3^{\nu\atop\bar\nu}\right],
\ee
where $G_F$ is the Fermi constant and $M_w$ the $W$ boson mass.  In the
parton model we have
\bee
  F_2^{\nu\atop\bar\nu}(x,Q^2) &=&
  \sum_q 2xf_q(x)+\sum_{\bar q}2xf_{\bar q}(x), \\
 xF_3^{\nu\atop\bar\nu}(x,Q^2) &=&
  \sum_q 2xf_q(x)-\sum_{\bar q}2xf_{\bar q}(x),
\eee
where the sums for neutrino scattering are over all partons that can
absorb a $W^+$, i.e., d, s, $\bar{\mathrm{u}}$ and $\bar{\mathrm{c}}$
and for antineutrino over those that can absorb a $W^-$, i.e., u, c,
$\bar{\mathrm{d}}$ and $\bar{\mathrm{s}}$.

\subsubsection{Global fits}

It is also possible to measure DIS on the neutron, or at least on
deuterium from which the neutron structure functions can be derived.
Using strong isospin symmetry, we have the relations
\bee
  f_{u/n}(x) &=& f_{d/p}(x), \\
  f_{\bar{u}/n}(x) &=& f_{\bar{d}/p}(x), \\
  f_{d/n}(x) &=& f_{u/p}(x), \\
  f_{s/n}(x) &=& f_{s/p}(x),
\eee
and so on.  It is conventional to always refer to the proton case,
dropping the ``$/p$'' subscript.  We therefore have the slightly
confusing result for $F_2^{en}$ shown below, in which $f_d$ is
multiplied by $(2/3)^2$, and so on.

We therefore have
\bee
  F_2^{ep} &=&
  \smfrac19xf_d+\smfrac49xf_u
  +\smfrac19xf_{\bar d}+\smfrac49xf_{\bar u}
  +\smfrac19xf_s+\smfrac19xf_{\bar s}
  +\smfrac49xf_c+\smfrac49xf_{\bar c}, \\
  F_2^{en} &=&
  \smfrac49xf_d+\smfrac19xf_u
  +\smfrac49xf_{\bar d}+\smfrac19xf_{\bar u}
  +\smfrac19xf_s+\smfrac19xf_{\bar s}
  +\smfrac49xf_c+\smfrac49xf_{\bar c}, \\
  F_2^{\nu p} &=&
  2xf_d+2xf_{\bar u}+2xf_s+2xf_{\bar c}, \\
 xF_3^{\nu p} &=&
  2xf_d-2xf_{\bar u}+2xf_s-2xf_{\bar c}, \\
  F_2^{\bar\nu p} &=&
  2xf_u+2xf_{\bar d}+2xf_c+2xf_{\bar s}, \\
 xF_3^{\bar\nu p} &=&
  2xf_u-2xf_{\bar d}+2xf_c-2xf_{\bar s}.
\eee
If we make the assumption that $f_{\bar s}=f_s$ and $f_{\bar c}=f_c$,
then we have six unknowns for six pieces of data so, given precise
enough data, we could solve for all the pdfs exactly.  In practice of
course it is never so simple and one must make global fits to as wide a
variety of data as possible.

\begin{figure}[t]
\vspace{.9cm}
  \begin{center}
  \rotatebox{-90}{\resizebox{8cm}{!}{\includegraphics{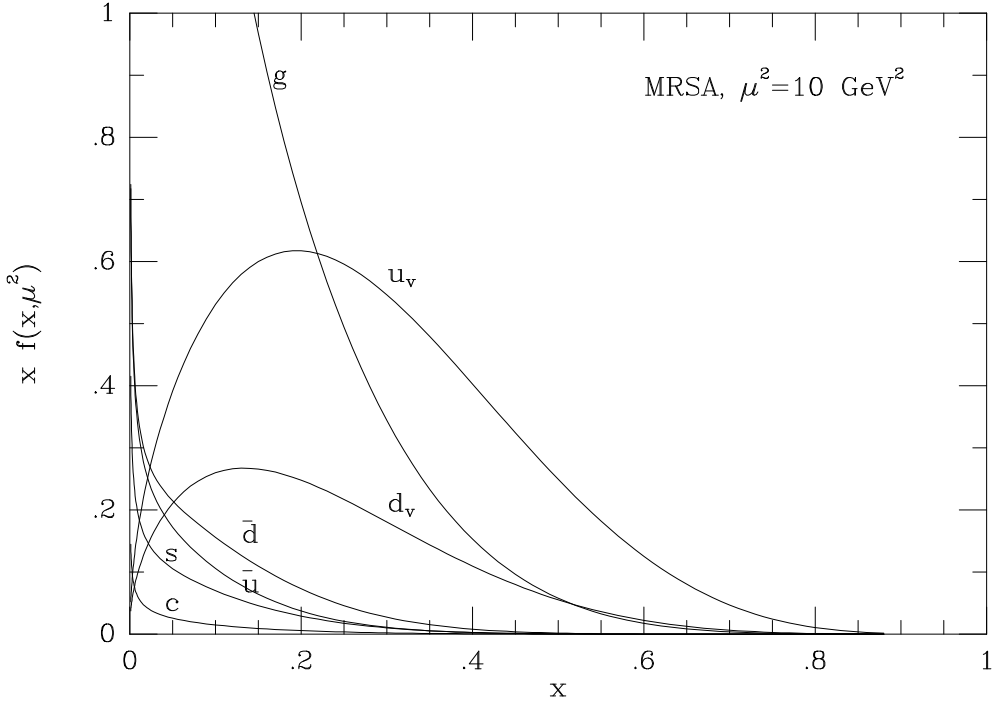}}}
  \end{center}
  \caption{\label{MRSA}Parton distribution function set A from the
    Martin-Roberts-Stirling group, taken from ESW\cite{ESW}}
\end{figure}

One gets typical results like those shown in Fig.~\ref{MRSA}.
Note that this uses the common notation of defining valence quark
distributions,
\bee
  f_{u_v} &\equiv& f_u-f_{\bar u}, \\
  f_{d_v} &\equiv& f_d-f_{\bar d}.
\eee
Non-valence quarks are generically referred to as the sea.

\subsubsection{Sum rules}

Having results for the pdfs, one can form interesting integrals over
them, for example,
\bee
  \int_0^1 dx \, f_{u_v}(x) &=& 2, \\
  \int_0^1 dx \, f_{d_v}(x) &=& 1.
\eee
Various such integrals can be constructed directly from the structure
functions.  It is worth checking that you can reproduce the physical
interpretation of each.

\paragraph{The Gross--Llewellyn-Smith sum rule}
\be
  \smfrac12\int_0^1 dx \left(F_3^{\nu p}+F_3^{\bar\nu p}\right)=3,
\ee
which counts the number of valence quarks in the proton.  In QCD this
provides a useful measurement of $\as$, because the right-hand side is
actually equal to $3\left(1-\frac{\as}{\pi}+{\cal O}(\as^2)\right)$.

\paragraph{The Adler sum rule}
\be
  \smfrac12\int_0^1 \frac{dx}x\left(F_2^{\bar\nu p}-F_2^{\nu p}\right)=1,
\ee
which counts the difference between the number of up and down valence
quarks.  This has the property that it is exact even in QCD, i.e., all
higher order corrections vanish.

\paragraph{The Gottfried sum rule}
\be
  \int_0^1 \frac{dx}x\left(F_2^{ep}-F_2^{en}\right)\approx0.23,
\ee
where the result is experimental.  This is sensitive to the difference
between the number of up and down sea quarks: it would be 1/3 if they
were equal.

\paragraph{The momentum sum rule}
Finally, we have the particularly significant result
\be
  \smfrac12 \int_0^1 dx \left(F_2^{\nu p}+F_2^{\bar\nu p}\right)
  \approx0.5,
\ee
where the result is again experimental.  This tells us that only about
half of the proton's momentum is carried by quarks and antiquarks.

\subsection{Hadronic collisions}

\subsubsection{The Drell--Yan process}

If the parton model is correct, the parton distribution functions should
be universal.  We should therefore be able to use the DIS measurements
to make predictions for other hadronic scattering processes.  The
classic example is the so-called Drell--Yan process, of lepton pair
production in hadron collisions,
\be
  h_1+h_2 \to \mu^++\mu^-+X,
\ee
where the state $X$ goes unmeasured.  In the parton model this arises as
the sum over all quark types of
\be
  q+\bar q \to \mu^++\mu^-.
\ee
The cross section can be written as the convolution of pdfs with a
partonic cross section, exactly like in DIS:
\be
  \hspace*{-1cm}
  \frac{d\sigma(h_1(p_1)+h_2(p_2)\to\mu^+\mu^-)}{dM^2} =
  \sum_q \int_0^1 d\eta_1 f_{q/h_1}(\eta_1)
  \int_0^1 d\eta_2 f_{\bar{q}/h_2}(\eta_2)
  \frac{d\sigma(q(\eta_1p_1)+\bar q(\eta_2p_2)\to\mu^+\mu^-)}{dM^2},
  \hspace*{-1cm}
\ee
where $M$ is the mass of the $\mu^+\mu^-$ pair.  Note that since the
partonic cross section contains a $\delta(M^2-\eta_1\eta_2s)$ term,
binning the data in $M$ gives extra information about the pdfs.  In
fact, binning also in the rapidity of the lepton pair, defined by
\be
  y \equiv \frac12\ln\frac
  {E_{\mu^+\mu^-}+p_{z,\mu^+\mu^-}}{E_{\mu^+\mu^-}-p_{z,\mu^+\mu^-}},
\ee
both $\eta$ values are fixed, providing a direct measurement of the
parton distribution functions (the partonic cross section can easily be
obtained by crossing the $e^+e^-\to q\bar{q}$ one we calculated in
Section~\ref{sec:eeqq}, divided by a factor of $N_c^2$ for the average
over incoming colours):
\be
  \frac{d^2\sigma}{dM^2dy} = \frac{4\pi\alpha^2}{3N_cM^2s}
  \sum_q e_q^2 f_{q/h_1}(e^y M/\surd s)
  f_{\bar{q}/h_2}(e^{-y} M/\surd s).
\ee

Note that the case $h_1=h_2=p$ provides a particularly good measure of
the sea quark distribution functions, which are hard to extract from
DIS data.

\subsubsection{Prompt photon and jet production}

Although we have not yet mentioned gluons, we will see in the next
lecture that there is also a non-zero pdf for the gluon, $f_g(\eta)$, as
can also be inferred from the momentum sum rule mentioned earlier.  As
well as being important for higher order corrections to the processes
given above, there are many processes in which they participate at tree
level.  The most important of these are prompt photon production,
\be
  h_1 + h_2 \to \gamma + X,
\ee
and jet production
\bee
  h_1 + h_2 &\to& q + q + X, \\
  h_1 + h_2 &\to& q + \bar q + X, \\
  h_1 + h_2 &\to& q + g + X, \\
  h_1 + h_2 &\to& g + g + X, \mbox{~etc.}
\eee

The gluon pdf is used in exactly the same way as the quark ones, and
hadronic cross sections can still be calculated as the sum of
convolutions of pdfs with partonic cross sections.  Prompt photon
production receives contributions from two partonic processes,
\bee
  \label{qqgamma}
  q + \bar{q} &\to& \gamma + g, \\
  \label{qggamma}
  q + g &\to& \gamma+q.
\eee
In the case $h_1=h_2=p$, the latter dominates, providing a measure of
the gluon pdf.  However there is a slight complication, in that
processes (\ref{qqgamma}), (\ref{qggamma}) are proportional to $\as$,
which is less well-known than $\alpha$, which controls the other
processes we have studied.  In fact this is always the case, that
measurements of the gluon pdf actually measure $\as\times f_g$ in
general.  The QCD corrections to this process turn out to be a lot
larger than any of the others we have considered, further complicating
this measurement.

\subsection{Summary}

We have considered the tree-level phenomenology of $e^+e^-$
annihilation, deep inelastic scattering and, more briefly, hadron
collisions.  It is remarkable how much QCD phenomenology can be
understood using tree level results.  However, we have to worry that
$\as$ is not so small, so higher order corrections must be important.
Equally importantly, it would be nice to see whether, and if so how, the
parton model emerges from QCD.

We discuss both these issues in the next lecture.

\section{Higher order corrections}
\setcounter{equation}{0}
\setcounter{footnote}{0}
\setcounter{figure}{0}

\subsection[$e^+e^-$ annihilation at one loop]
 {\boldmath $e^+e^-$ annihilation at one loop}

In this section, I go through the calculation of the NLO correction to
the $e^+e^-\to$ hadrons cross section in some detail.  I will briefly
describe some of the more technical aspects of the calculation, for
those interested, in Section~\ref{canskip}, but those who are not can
safely skip this section, since I recap the important results at the
start of Section~\ref{canskip+1}.

In discussing the $e^+e^-\to$ hadrons cross section at tree level, we
assumed that quarks produce hadrons with probability~1.  Therefore we
calculated the $e^+e^-\to q\bar{q}$ cross section in
Section~\ref{sec:eeqq}.  In discussing jet cross sections, we extended
this to say that all partons produce hadrons with probability~1.
Therefore we should calculate the total cross section to produce any
number or type of partons.  At leading order this makes no difference,
since the only possible process is $e^+e^-\to q\bar{q}$, but at order
$\as$ we have to calculate and sum the cross sections for $q\bar q$ and
$q\bar qg$ final states.  We start with the latter.

Recall that the total $q\bar qg$ cross section is divergent,
\be
  \sigma = \sigma_0\;
  C_F\frac{\alpha_s}{2\pi}
  \int dx_1\,dx_2\frac{x_1^2+x_2^2}{(1-x_1)(1-x_2)},
\ee
where the region of integration is the upper right triangle of the unit
square, bordered by the lines $x_1=1$ and $x_2=1$, which are the
singular regions.  This divergence must be regularized in some way,
before we can make progress.

First though we discuss the origin of the divergences.  They arise from
propagator factors that diverge,
\be
  \raisebox{-5ex}{\scalebox{0.7}{%
      \includegraphics*[64,374][262,446]{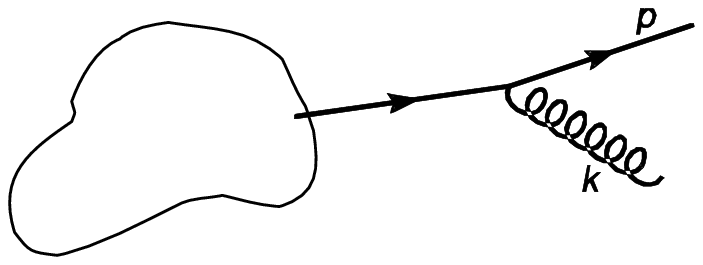}}}
  \qquad
  \frac1{(p+k)^2}=\frac1{2p\ldot k}=
\frac1{2E\omega(1-\cos\theta)}
\approx\frac1{E\omega\theta^2},
\ee
where $E$ and $\omega$ are the quark and gluon energies and $\theta$ is
the angle between them.

In the collinear limit, $\theta\to0$, one in principle obtains
$1/\theta^4$ in the matrix element squared, but in fact the numerators
always contribute a factor of $\theta^2$, so one obtains
\be
  |\mathcal{M}|^2 \sim \frac1{\theta^2}.
\ee
In the soft limit, $\omega\to0$, one has in the interference between
diagrams in which the gluon is attached to quark~1 and quark~2,
\be
|\mathcal{M}|^2\sim\frac{p_1\ldot p_2}{p_1\ldot k\,p_2\ldot k}
\sim\frac1{\omega^2}.
\ee
In terms of $\omega$ and $\theta$ the phase space is given by
\be
\frac{d^3k}{2\omega}=\smfrac12\,
\omega d\omega\,\sin\theta\,d\theta\,d\phi
\sim\omega d\omega\,\theta d\theta.
\ee
We therefore have logarithmic singularities in both the soft and
collinear limits.  We generically refer to both of these as the infrared
limit.

\subsubsection{Regularization}

As in the discussion of renormalization, the simplest way we could
regularize this cross section is with a cutoff, for example on the
transverse momentum of the gluon, which would prevent the integration
entering both the soft and collinear regions.  However, we will see that
infrared singularities cancel between different contributions, in this
case $q\bar q$ and $q\bar qg$, so we must use a regularization that can
be consistently applied in all contributions.  It is not clear that this
is the case for a cutoff, since it must be applied in both real and
virtual contributions, which have very different structures.  Instead,
to ensure consistent application across all processes, it is better to
modify the theory in such a way that some dimensionless parameter
$\epsilon$ regulates the divergences.  Then the complete calculation can
be performed in this modified theory and at the end of the calculation,
when all the divergences have cancelled, the limit $\epsilon\to0$ can be
smoothly taken.  Remarkably, dimensional regularization, which we used
for ultraviolet singularities, also provides a consistent regulator for
infrared singularities, as we shall discuss in detail shortly.

Another regularization scheme, which actually works well in QED, and for
simple processes in QCD, is the gluon (or photon) mass regularization.
We introduce a non-zero gluon mass $m_g^2=\epsilon Q^2$.  This prevents
the propagators from reaching zero and diverging: for massless quarks
the minimum value is $m_g^2$ and for a quark of mass $m_q$ it is
$2m_qm_g$.  With this modification one can recalculate the differential
cross section and integrate it to give a finite result,
\be
    \sigma_{q\bar{q}g} = \sigma_0 \; C_F\frac{\alpha_s}{2\pi}
    \left(
      \log^2\frac1\epsilon - 3\log\frac1\epsilon + 7 - \frac{\pi^2}3
      +{\cal O}(\epsilon)\right).
\ee
However, since a non-zero gluon mass violates gauge invariance, this
method is bound to fail in general.  In particular, it is not suitable
for any process in which any lowest order contributions have external
gluons.  As in the ultraviolet case, the only scheme that is known to be
consistent with all the symmetries of QCD, and hence to work to
arbitrary orders in arbitrary processes, is dimensional regularization.

The reason why I said that it is remarkable that dimensional
regularization works in the infrared limit is the fact that the two
limits have non-overlapping regions of applicability in the complex $d$
plane.  Ultraviolet-singular integrals are regularized by working in
$d<4$ dimensions, but infrared-singular integrals are only rendered
finite by working in $d>4$ dimensions.  However, by carefully splitting
contributions that are singular in both the infrared and ultraviolet one
can consider the regularization schemes that are used in each as
independent.  In each region, one considers the appropriate
dimensionality ($d=4\!-2\epsilon$ with $\epsilon>0$ in the ultraviolet
and with $\epsilon<0$ in the infrared) and then analytically continues
to the whole complex $\epsilon$ plane.  Since analytical continuation is
unique, this gives a unique result for each, in the region of
applicability of the other, and the two can be combined before the limit
$\epsilon\to0$ is taken.  This subtlety leads to some surprising
results, for example for the self-energy of a massless quark, discussed
below.

As the calculation of cross sections in dimensional regularization is
rather technical, it is rare to see it done in summer school lectures,
but I think it brings out some interesting points, so I at least sketch
how the calculation works in Section~\ref{canskip}.  As I said, those
who disagree can safely skip ahead to Section~\ref{canskip+1}.

\subsubsection{Aside: Real and virtual corrections in dimensional
  regularization}
\label{canskip}

It is straightforward to generalize the Feynman rules to $d$ dimensions
and fairly straightforward to generalize the Dirac algebra.  The result
is that $d$-dimensional matrix elements still have propagators
$\sim1/p^2$, but that the numerators become $d$~dependent.  (It is worth
mentioning the closely-related dimensional reduction scheme, which is
often used for supersymmetry calculations, since conventional
dimensional regularization violates supersymmetry.  In this scheme one
works in $d$ dimensions, but modifies the theory in such a way that
fermions and massless vector bosons still have 2 spin states, instead of
$d\!-\!2$ as in dimensional regularization.  The result is that the
matrix elements themselves are equal to the 4-dimensional ones and it is
only on performing the loop and phase space integrals that the $d$
dimensionality gets introduced.)

\paragraph{Phase space integrals}
We will have to integrate over $d$-dimensional phase space.  We begin by
considering integer values of $d$ and then continue the results to real
values.  It is straightforward to write down the basic integration
measure,
\be
  d^dk\,\delta_+(k^2) = \frac{d^{d-1}k}{2\omega}
  = \frac12 \omega^{d-3}d\omega\,d\Omega_{d-2},
\ee
where $\omega$ is the energy of $k$ and $d\Omega_{d-2}$ is an element of
$d\!-\!2$-dimensional solid angle.  The only difficulty concerns the
evaluation of integrals over this solid angle.  In four dimensions we
have
\be
  k = \omega(1;\sin\phi\,\sin\theta,\cos\phi\,\sin\theta,\cos\theta)
  \qquad\mbox{(4 dimensions)},
\ee
where $\theta$ and $\phi$ are the usual spherical polar coordinates with
$\theta$ the polar angle and $\phi$ the azimuthal angle.  In five
dimensions we have
\be
  k = \omega(1;\sin\psi\,\sin\phi\,\sin\theta,\cos\psi\,\sin\phi\,\sin\theta,
  \cos\phi\,\sin\theta,\cos\theta)
  \qquad\mbox{(5 dimensions)},
\ee
where $\psi$ is an azimuthal angle in the additional dimension.
Generalizing to $d$ dimensions, we have $d-4$ additional azimuths and we
write $k$ generically as
\be
  k = \omega(1;\ldots,\cos\phi\,\sin\theta,\cos\theta)
  \qquad\mbox{($d$ dimensions)},
\ee
where the ellipsis represents a $d\!-\!3$-vector of length
$\sin\phi\,\sin\theta$ containing $d-4$ azimuths.  Depending on the
complexity of the calculation, more or less of these
additional components have to be specified precisely.  In fact in our
case, since we only consider the relative orientations of three momenta
that have zero total momentum, and therefore all lie in a plane, it is
sufficient to specify
\be
  k = \omega(1;\ldots,\cos\theta)
  \qquad\mbox{($d$ dimensions)},
\ee
where the ellipsis represents a $d\!-\!2$-vector of length
$\sin\theta$ containing $d-3$ azimuths.

We can see how to integrate over the additional azimuths by again
considering integer $d$ and then generalizing,
\bee
  \int d\Omega_1 &=& \int d\phi = 2\pi, \\
  \int d\Omega_2 &=& \int d\phi \; \sin\theta \, d\theta = 4\pi, \\
  \int d\Omega_3 &=& \int d\psi \; \sin\phi\, d\phi \; \sin^2\theta \,
  d\theta = 2\pi^2,
\eee
and so on.  We have a recursion relation
\be
  \int d\Omega_n = \int d\Omega_{n-1} \; \sin^{n-1}\theta\, d\theta,
\ee
which is solved by
\be
  \Omega_n \equiv \int d\Omega_n =
  \frac{2\pi^{(n+1)/2}}{\Gamma[(n+1)/2]}.
\ee
We are now equipped to tackle the phase space integral, and see how the
dimensional regularization succeeds in regularizing our integrals.

\paragraph{Regularization}  Since the form of the propagator factors is
unchanged in $d$ dimensions, and it is these that dominate the singular
region, it is straightforward to read off the behaviour in the
regularized theory.  In the soft region we have
\be
\label{eq:softreg}
\int_0\omega^{1-2\epsilon}d\omega\;\frac1{\omega^2}
=\int_0\frac{d\omega}{\omega^{1+2\epsilon}}
\sim-\frac1{2\epsilon},
\qquad\epsilon<0,
\ee
and in the collinear region
\be
\label{eq:collreg}
\int_0\sin^{1-2\epsilon}\theta\,d\theta\;\frac1{\theta^2}
\sim\int_0\frac{d\theta}{\theta^{1-2\epsilon}}
\sim-\frac1{2\epsilon},
\qquad\epsilon<0.
\ee
Since our cross section is divergent in both limits, and they can
overlap, i.e., a radiated gluon can be both soft and collinear, we expect
the total cross section to be of order $1/\epsilon^2$.  Note, as a
consistency check, that the integrands are positive definite and that,
in the region in which they are well-defined, $\epsilon<0$, the results
are positive (and divergent as $\epsilon\to0$).

\paragraph{Total \boldmath$e^+e^-\to q\bar qg$ cross section}
We now have all the ingredients we need to calculate the differential
cross section for $e^+e^-\to q\bar qg$ and to integrate it over all
phase space in dimensional regularization.  We obtain
\be
\label{eq:sigmaqqg}
    \sigma_{q\bar{q}g} = \sigma_0 \; C_F\frac{\alpha_s}{2\pi}
    H(\epsilon)
    \left(
      \frac2{\epsilon^2} + \frac3\epsilon + \frac{19}2-\pi^2
      +{\cal O}(\epsilon)\right),
\ee
where $\sigma_0$ is the lowest order cross section and $H(\epsilon)$ is
a smooth function, with $H(0)=1$, that we will not ultimately need to
know.  Note that, as we anticipated from
Eqs.~(\ref{eq:softreg}) and~(\ref{eq:collreg}), this result is positive, and
divergent like $1/\epsilon^2$ as $\epsilon\to0$.

So far, the regularization scheme has succeeded in quantifying the
degree of divergence of the total three-parton cross section, but it has
not helped us solve the problem of the divergence, by recovering a
finite result for a physical cross section.  As we already anticipated
above, this will come by calculating the loop correction to $e^+e^-\to
q\bar q$.

\begin{figure}[t]
  \centerline{%
    \includegraphics{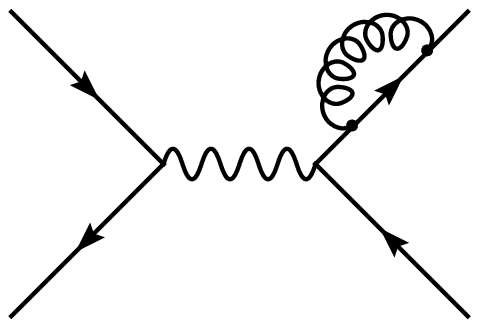}\hfill
    \includegraphics{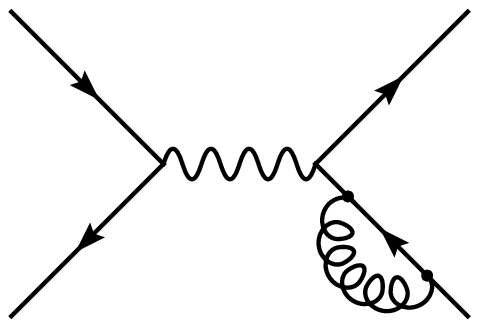}\hfill
    \includegraphics{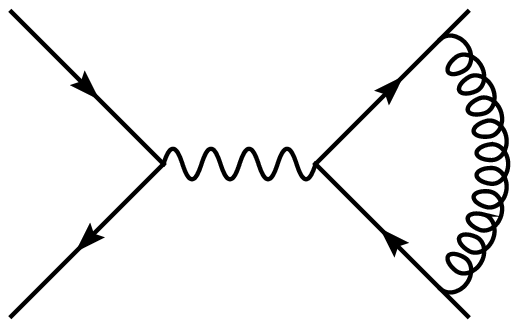}}
  \caption{One-loop diagrams for $e^+e^-\to q\bar q$}
  \label{fig:eeloop}
\end{figure}

\paragraph{\boldmath$\sigma(e^+e^-\to q\bar q)$ at one loop}
We already made the point that to calculate the total cross section for
$e^+e^-\to$ hadrons, we must sum over all $e^+e^-\to$ partons processes.
At this order of perturbation theory $q\bar q$ is the only other process
that contributes.  There are three diagrams, shown in
Fig.~\ref{fig:eeloop}.
They are down by one power of $\as$ relative to the tree-level diagram,
\be
  \mathcal{M}_1 \propto \as \mathcal{M}_0.
\ee
Therefore $|\mathcal{M}_1|^2$ is two powers down and hence negligible at
the order to which we are working.  However, since the final state is
the same as that of the tree-level diagram, the two interfere, and their
interference, $\mathrm{I\!Re}\{\mathcal{M}_0^*\mathcal{M}_1\}$ does
contribute at order $\as$.

In quantum mechanics, you know that we must sum over all unobserved
quantum numbers at the amplitude level.  Since the gluon momentum is
unconstrained by the outgoing quark momenta, we must sum over {\em all}
gluon momenta,
\be
  \int d^dk.
\ee
Note that there is no mass-shell-constraining delta-function: the
virtual integral is over all arbitrary on- and off-shell momenta.

We begin with the first two diagrams, which are proportional to the
self-energy of a massless quark.  It is actually easy to see that these
have to be zero in dimensional regularization: the value of the integral
has dimensions $E^{d-4}$, but by Lorentz invariance the result of the
integral can only be a function of the square of the quark's momentum,
$p^2=0$, so there is nothing that can provide this
dimensionality\footnote{In fact this statement relies on working in a
covariant gauge.  In a lightcone gauge for example, the self-energy can
depend on $n\ldot p$.  This diagram is not then zero, but of course the
final answer for the sum of the three diagrams is gauge invariant.}.
The only way these two facts can be reconciled is if the integral is
zero.  However, if we examine the integrand somewhat closer, this is
very surprising, because it is positive definite.  How can a positive
definite quantity integrate to zero?

The answer to this question comes from a subtle use of dimensional
regularization.  In fact this integral is divergent in both the infrared
and ultraviolet.  If we split the integral into two parts by introducing
an arbitrary separation scale $\Lambda$, then we obtain an ultraviolet
contribution $\sim\Lambda^{-2\epsilon}/\epsilon$ and an infrared
contribution $\sim-\Lambda^{-2\epsilon}/\epsilon$.  Each is positive in
its domain of applicability ($\epsilon>0$ and $\epsilon<0$
respectively), but after analytically continuing each to arbitrary
$\epsilon$, they are exactly equal and opposite, giving a zero result
for these diagrams.

Turning to the third diagram, the vertex correction, we find that it is
also divergent in the infrared and ultraviolet regions.  However, its
ultraviolet divergence is exactly equal and opposite to the one from the
sum of the two self-energy diagrams.  Therefore the sum of the three
diagrams is ultraviolet finite and no renormalization is needed at this
order.  This actually follows directly from the Ward identity of QED.
Thus, one simply has to evaluate the vertex correction diagram in
dimensional regularization, to obtain the complete order $\as$
contribution to $e^+e^-\to q\bar q$.  We find that the infrared
divergences do not cancel, and we obtain
\be
\label{eq:sigmaqq}
    \sigma_{q\bar{q}} = \sigma_0 \; C_F\frac{\alpha_s}{2\pi}
    H(\epsilon)
    \left(
      -\frac2{\epsilon^2} - \frac3\epsilon - 8+\pi^2
      +{\cal O}(\epsilon)\right).
\ee
Dimensional regularization has succeeded in regularizing the divergence
of this contribution as well.  This time, however, the result is negative
and divergent as $\epsilon\to0$.  This should not surprise us, as we
already noted that this is an interference term, so there is no
requirement that it be positive, as there was for $\sigma_{q\bar{q}g}$.
In fact a quick glance at Eqs.~(\ref{eq:sigmaqqg}) and~\ref{eq:sigmaqq})
shows us that the divergences are going to cancel between them.

\subsubsection{The total cross section}
\label{canskip+1}

In the previous section we discussed how dimensional regularization
provides finite results for the total cross sections for the $e^+e^-\to
q\bar q$ and $e^+e^-\to q\bar qg$ processes, which each diverge as
$\epsilon\to0$.  For the benefit of those who slept through it, I restate
them here:
\bee
    \sigma_{q\bar{q}} &=& \sigma_0 \; C_F\frac{\alpha_s}{2\pi}
    H(\epsilon)
    \left(
      -\frac2{\epsilon^2} - \frac3\epsilon - 8+\pi^2
      +{\cal O}(\epsilon)\right), \\
    \sigma_{q\bar{q}g} &=& \sigma_0 \; C_F\frac{\alpha_s}{2\pi}
    H(\epsilon)
    \left(
      \phantom{-}\frac2{\epsilon^2} + \frac3\epsilon + \frac{19}2 -\pi^2
      +{\cal O}(\epsilon)\right).
\eee
According to our earlier discussion, the total cross section for
$e^+e^-\to$ hadrons is given by the sum of the two.  It is finite, so
the limit $\epsilon\to0$ can be taken,
\pagebreak[3]%
\bee
\sigma_{e^+e^-\to\mbox{\small hadrons}}
&=&\sigma_0\left(
1+C_F\frac{\as}{2\pi}\,\frac32\right)\\
&=&\sigma_0\left(
1+\frac{\as}{\pi}\right).
\label{eq:aspi}
\eee
Of course, this would be useless if it depended on the regularization
procedure.  The proof of its independence is beyond us here, but it is
worth demonstrating it, by comparison with another scheme, the gluon
mass regularization, in which we have
\bee
  \sigma_{q\bar q} &=& \sigma_0\,C_F\frac{\as}{2\pi}
  \left[-\log^2\frac1\epsilon+3\log\frac1\epsilon
  -\frac{11}2+\frac{\pi^2}3
  +{\cal O}\left(\epsilon\right)\right], \\
  \sigma_{q\bar qg} &=& \sigma_0\,C_F\frac{\as}{2\pi}
  \left[\phantom{-}\log^2\frac1\epsilon-3\log\frac1\epsilon
  +7-\frac{\pi^2}3
  +{\cal O}\left(\epsilon\right)\right], \\
  \sigma_{\mathrm{had}} &=& \sigma_0
  \left[1+\frac{\as}{\pi}\right].
\eee
Note that the individual cross sections have completely different forms
in the different schemes, but that the sum of the two is scheme
independent.

Equation~(\ref{eq:aspi}) is one of the most fundamental quantities in
QCD and is certainly one of the most well-calculated and measured.
Despite the fact that it is a relative small correction to the total
rate, experimental and theoretical systematic errors are so small that
they can almost be neglected~--- even with the large statistics of $\tau$
decays and $Z$ decays at LEP, the statistical errors dominate.  This
means that not only does it provide one of the most accurate
measurements, but its quoted accuracy is rather easy to interpret and
implement in global analyses for example, unlike measurements that are
dominated by systematics.

\begin{figure}[t]
  \centerline{\includegraphics*[225,180][521,386]{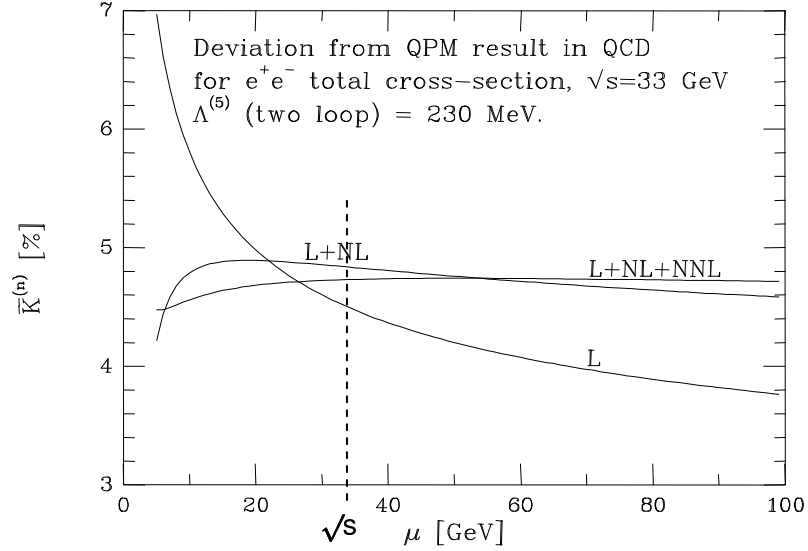}}
  \caption{The QCD prediction for the corrections to $R_{e^+e^-}$ at
  $\surd s=33$~GeV as a function of renormalization scale at leading,
  next-to-leading, and next-to-next-to-leading order, taken from
  ESW~\cite{ESW}}
  \label{fig:repemho}
\end{figure}

Equation~(\ref{eq:aspi}) is now known up to order $\as^3$.  As discussed
in Section~\ref{renormalization}, renormalization introduces a
renormalization scale dependence into $\alpha_s$ and the coefficient
functions beyond the first one,
\be
\sigma_{e^+e^-\to\mathrm{hadrons}}
=\sigma_0\left(
1+\frac{\as(\mu)}{\pi}+
C_2\!\left(\frac{\mu^2}s\right)\left(\frac{\as(\mu)}{\pi}\right)^2+
C_3\!\left(\frac{\mu^2}s\right)\left(\frac{\as(\mu)}{\pi}\right)^3\right).
\ee
Reducing this renormalization-scale dependence is one of the biggest
reasons for going to higher orders.  As can be seen in
Fig.~\ref{fig:repemho}, the scale-dependence is indeed significantly
smaller at each order, giving stability over a wider range of $\mu$.
It can also be seen that provided $\mu$ is of order $Q$, the higher
order corrections are relatively small.  We will see shortly that simply
taking the leading order result with $\mu=\surd s$ does surprisingly
well and is certainly sufficient to understand the phenomenology.

\subsubsection{$\as$ measurements}
\label{sec:asmeasure}

As I mentioned above, the experimental measurement of $R_{e^+e^-}$ gives
one of the best measurements of~$\alpha_s$.  In fact the LEP combined
value of $R_{had}$ is
\be
  R(LEP) = 20.767\pm0.025,
\ee
while the tree-level prediction is
\be
  R_0(M_z) = 19.984.
\ee
Combining the two, and simply using the leading order result with
$\mu=M_z$, we obtain our first measurement of $\as$,
\be
  \as(M_z) = 0.124\pm0.004,
\ee
surprisingly close to the value using the four-loop
result~\cite{Davier:2008sk}, $0.119\pm0.003$.

As we discussed in Section~\ref{sec:measureas}, since QCD predicts the
scale dependence of $\as$, one measurement at any scale is sufficient to
give a prediction for all scales.  We can therefore phrase measurements
at other scales either as tests of QCD throughout the intervening energy
range or, by translating them all into measurements at a single scale,
as different measurements of the same quantity that can be combined to
give a better overall measurement.

As an example, the average measurement of $R$ over several energy points
around 34~GeV is
\be
  R(PETRA) = 3.88\pm0.03,
\ee
while the tree-level prediction is
\be
  R_0(\mbox{34 GeV}) = 3.69.
\ee
Again using the leading order result, we obtain
\be
  \as(\mbox{34 GeV}) = 0.162\pm0.026.
\ee
Finally, using the one-loop renormalization group equation, we can
convert this into a measurement of $\as(M_z)$,
\be
  \as(M_z) = 0.134\pm0.018.
\ee
This is in good agreement with the value from LEP, although with much
larger uncertainties, simply due to the fact that the statistics of the
PETRA experiments were much lower.

As a final example, we consider $\tau$ decays.  The QCD corrections to
the hadronic decay rate actually have two effects: on the ratio of
branching fractions, $R_\tau$, as discussed earlier, and also directly
on the total decay rate of the $\tau$.  These can form the basis for two
analyses in which the experimental errors are largely independent.  The
combined result for the two is
\be
  \as(M_\tau=\mbox{1.77 GeV}) = 0.34\pm0.01.
\ee
This time, because we are translating over such a wide energy range the
one-loop renormalization group equation does not do quite such a good
job,
\be
  \as^{\mbox{\small(one-loop)}}(M_z)=0.1272\pm0.0014,
\ee
compared to the four-loop value~\cite{Davier:2008sk}
\be
  \as(M_z)=0.1212\pm0.0011,
\ee
but it is not so far out.  Note in this case the phenomenon of the
`incredible shrinking error'.  Although the measurement at the $\tau$
mass scale has a precision of about 3\%, after evolving it to $M_z$ the
relative uncertainty gets scaled down by the ratio of the two $\alpha_s$
values, and $\tau$ decays give the best single measurement of
$\alpha_s(M_z)$.

\begin{figure}[t]
  \centerline{%
  \scalebox{0.7}{\includegraphics{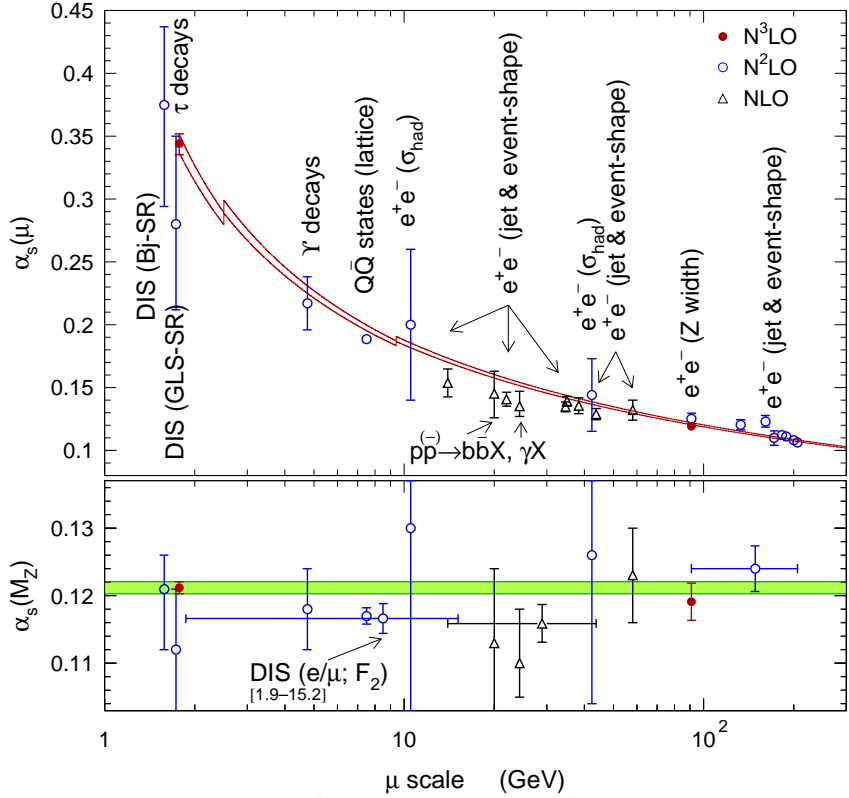}}}
  \vspace*{-1cm}
  \caption{Results of a recent compilation of $\as$
    values~\cite{Davier:2008sk,Bethke:2006ac}}
  \label{fig:sigge}
\end{figure}

The results of a recent compilation~\cite{Davier:2008sk,Bethke:2006ac}
are shown in Fig.~\ref{fig:sigge}.
The scale dependence shows excellent agreement with the predictions of
perturbative QCD over a wide energy range.  When translated into
measurements of $\as(M_z)$, the separate measurements cluster strongly
around the average value,
\be
  \as^{\mbox{\small(average)}}(M_z) = 0.1204\pm0.0009.
\ee


\subsection{Deep inelastic scattering revisited}

The parton model I described in the last lecture assumed that the
partons are non-interacting.  But we know that they do interact via QCD,
so what will happen when we consider these interactions?  We will
discover that the structure functions do become slowly (logarithmically)
varying with $Q^2$.  We start by considering the next-to-leading order
QCD corrections to quark scattering.  We will find that these, if
calculated naively, would be divergent, but that these divergences can
be absorbed into the parton distribution functions.  These will then
become scale-dependent, giving rise to the $Q^2$-dependence of the
structure functions.

\subsubsection{NLO corrections to DIS}

The next-to-leading order corrections come from three sources (recalling
that we sum and integrate over all final states $X$, so we must sum over
all contributions in which any kind of parton is scattered):
\begin{enumerate}
\item One-loop corrections to $eq \to eq$,
\item $eq \to eqg$,
\item $eg \to eq\bar{q}$.
\end{enumerate}
The~third contribution is completely new in QCD and is not present in
the parton model.  We come back to it in a later section.  The other two
can more genuinely be thought of as higher-order corrections to the
parton model process.  We start with the~second.

There are two contributing diagrams.  The matrix element squared can
be obtained by crossing from $e^+e^- \to q\bar{q}g$ (\ref{qqgfinal}).
Labeling the momenta as
\be
  e(k) + q(\eta p) \to e(k') + q(p_1) + g(p_2),
\ee
we obtain
\be
  \sum|{\cal M}|^2 = \frac{8C_FN_ce^4e_q^2g_s^2}
  {k\ldot k'\,p_1\ldot p_2\,\eta p\ldot p_2}
  \left((p_1\ldot k)^2+(\eta p\ldot k)^2
       +(p_1\ldot k')^2+(\eta p\ldot k')^2\right).
\ee
As usual the phase space is given by (\ref{PSfact}),
\be
  dPS = \frac{Q^2}{16\pi^2 s x^2}dQ^2 \, dx \, dPS_X.
\ee
This time $X$ consists of two partons so is non-trivial,
\be
  dPS_X = \frac{d\cos\theta \,d\phi}{32\pi^2},
\ee
where $\theta$ and $\phi$ refer to the direction of $p_1$ in the
centre-of-mass system of $\eta p\!+\!q$.  It is conventional to replace
$\cos\theta$ by the manifestly Lorentz-invariant variable $z$,
\be
  z \equiv \frac{p_1\ldot p}{q\ldot p} = \smfrac12(1-\cos\theta),
\ee
with range $0<z<1$, giving
\be
  dPS_X = \frac{dz \,d\phi}{16\pi^2}.
\ee
It will later be instructive to know the transverse momentum of $p_1$ in
this frame,
\be
  k_\perp^2 = Q^2\left(\frac\eta x-1\right)z(1-z).
\ee
Note also that the case $\eta=x$ corresponds to a massless final state.
Kinematically this can only happen if either $p_1$ or $p_2$ are
infinitely soft (i.e., have zero energy), or if they are exactly
collinear.

We therefore have
\be
  \hspace*{-5mm}
  \frac{d\sigma^2(e+q)}{dx\,dQ^2} = \frac1{4N_c} \, \frac1{2\hat s}
  \, \frac{Q^2}{16\pi^2 s x^2} \int \frac{dz \,d\phi}{16\pi^2} \,
  \frac{8C_FN_ce^4e_q^2g_s^2}
  {k\ldot k'\,p_1\ldot p_2\,\eta p\ldot p_2}
  \left((p_1\ldot k)^2+(\eta p\ldot k)^2
       +(p_1\ldot k')^2+(\eta p\ldot k')^2\right).
  \hspace*{-5mm}
\ee
Rewriting in terms of our kinematic variables and averaging over $\phi$,
we have
\bee
  \lefteqn{
  \left\langle
  \frac{(p_1\ldot k)^2+(\eta p\ldot k)^2
       +(p_1\ldot k')^2+(\eta p\ldot k')^2}
  {k\ldot k'\,p_1\ldot p_2\,\eta p\ldot p_2}
  \right\rangle_\phi
}\nonumber\\&=&
  \frac1{y^2Q^2}\left(
    (1+(1-y)^2)\Biggl[\frac{1+x_p^2}{1-x_p}\,\frac{1+z^2}{1-z_{\phantom{p}}}
    +3-z-x_p+11x_pz\Biggr]
  -y^2\Biggl[8zx_p\Biggr]\right),
\phantom{(9.99)}
\eee
where $x_p=x/\eta$.  Two things are already clear: at this order we will
have a non-zero longitudinal structure function, $F_L(x,Q^2)$; and the
$z$ integration, which runs from 0~to~1, will give a divergent
contribution to $F_2$.  This should worry us, since we are calculating a
physical cross section, but let us continue for a while and see what
happens.

Putting everything together we have
\be
  \hspace*{-5mm}
  \frac{d\sigma^2(e+q)}{dx\,dQ^2} =
  \frac{C_F\alpha^2e_q^2\as}{2\eta x^2y^2s^2}
  \int_0^1 dz
  \left(
    (1+(1-y)^2)\Biggl[\frac{1+x_p^2}{1-x_p}\,\frac{1+z^2}{1-z_{\phantom{p}}}
    +3-z-x_p+11x_pz\Biggr]
  -y^2\Biggl[8zx_p\Biggr]\right),
  \hspace*{-5mm}
\ee
and hence
\be
  F_2(x,Q^2) =
  \sum_q\int_x^1 dx_p \, e_q^2\frac{x}{x_p}f_q\left(\frac{x}{x_p}\right) \,
  \frac{C_F\as}{2\pi}
  \int_0^1 dz
  \left(
    \frac{1+x_p^2}{1-x_p}\,\frac{1+z^2}{1-z_{\phantom{p}}}
    +3-z-x_p+11x_pz
  \right)\!.\!
\ee
The divergence at $z\to1$ corresponds to kinematic configurations in
which the outgoing gluon becomes exactly collinear with the incoming
quark.  Therefore in the Feynman diagram in which the gluon is attached
to the incoming quark, the internal quark line becomes on-shell, causing
the divergence.  Note also that the coefficient of the divergence itself
diverges at the point $x_p=1$, at which the gluon is infinitely soft.

In order to study the divergence, let us first regulate it by
calculating the contribution from emission with $k_\perp^2>\mu^2$
(and assume $\mu^2\ll Q^2$ for simplicity).  Since $k_\perp^2$ is
proportional to $(1\!-\!z)$ this will give us finite integrals.  At any
time, the full result can be obtained by setting $\mu\to0$.  We
therefore obtain
\be
  \label{Pxp}
  F_2(x,Q^2) =
  \sum_q\int_x^1 dx_p \, e_q^2\frac{x}{x_p}f_q\left(\frac{x}{x_p}\right) \,
  \frac{\as}{2\pi}
  \left(
    \hat P(x_p)\log\frac{Q^2}{\mu^2}
    +R(x_p)
  \right),
\ee
where the function $R(x_p)$ is finite.  In the following we will not
keep track of this function, although it would be essential for
quantitative analysis.  The function $P(x_p)$ we introduced in
(\ref{Pxp}) is called the splitting function (or more strictly speaking
the unregularized splitting function),
\be
  \hat P(x) = C_F\frac{1+x^2}{1-x_{\phantom{p}}}.
\ee
It actually describes the probability distribution of quarks produced in
a splitting process, $q\to qg$ in which the produced quark has a
fraction $x$ of the original quark's momentum.  (We will quantify this
statement slightly more shortly.)

Obviously by regulating the divergence we have not removed it: physical
cross sections are still supposed to be obtained by setting
$\mu\to0$, in which case $F_2$ is logarithmically divergent.
However, before discussing what happens to this divergence, let us
consider the virtual one-loop correction to $eq\to eq$.  Since this
diagram contains two quark-gluon couplings, when squared it would give
an ${\cal O}(\as^2)$ correction.  However, since it has the same final
state as the lowest order diagram, we must consider the interference
between the two, and this interference is ${\cal O}(\as)$, so we must
include it.

We could obtain the result for the one-loop diagram by crossing from
$e^+e^-\to q\bar q$.  However, to illustrate the physics, it is
sufficient to recall a few of its features.  Firstly, since the external
particles are the same as in the lowest-order process, the kinematics
must be the same.  In particular, it can only contribute at the
point~$\eta=x$.  Secondly, as in the $e^+e^-$ case, the interference of
the one-loop and tree-level diagrams is divergent and negative.  In fact
the kinematic regions in which the one-loop integrand diverges are
exactly the same as those of the $eq\to eqg$ contribution we have just
considered: when the gluon is soft, or is collinear with either of the
quarks.

It turns out that the divergence is exactly right to cancel the one we
obtained above at $x_p\to1$.  In fact one finds that after including the
one-loop contribution, one gets exactly the same formula as (\ref{Pxp})
except that the unregularized splitting function $\hat P(x_p)$ is
replaced by the regularized one, $P(x_p)$,
\be
  P(x) = \hat P(x) + P_{\mathrm{virtual}}(x).
\ee
Since the one-loop contribution has the same kinematics as the
lowest-order process, $P_{\mathrm{virtual}}(x)$ must be proportional to
$\delta(1-x)$.  $P(x)$ is therefore a distribution.

To define it, we will need to use a mathematical trick called the
plus-distribution.  Given some function $f(x)$, which is well-defined
for all $0\le x<1$, we define a distribution $f(x)_+$ on the region
$0\le x\le1$, as
\be
  f(x)_+ = f(x) - \delta(1-x)\int_0^1 dx'f(x').
\ee
The plus-distribution is most useful when the function $f(x)$ is
divergent at $x\to1$.  This means that for any other function $g(x)$,
which is smooth at $x=1$, we have the property
\be
  \int_0^1 dx\,f(x)_+\,g(x) = \int_0^1 dx\,f(x)\left(g(x)-g(1)\right).
\ee
Provided that $g(x)$ goes to $g(1)$ sufficiently quickly, this integral
is finite.

After including the virtual contribution, the splitting function is
given by
\be
  P(x) = C_F\left[\frac{1+x^2}{(1-x)_+}+\frac32\delta(1-x)\right].
\ee
This is actually the first correction to a function that describes the
momentum distribution of quarks within quarks,
\be
  {\cal P}(x) \equiv \delta(1-x) +
  \frac{\as}{2\pi}\log\frac{Q^2}{Q_0^2}P(x)
+ {\cal O}(\as^2\log^2),
\ee
where the distribution is defined to be a pure quark at scale $Q_0$ and
probed at scale $Q$.

Inserting the full splitting function into (\ref{Pxp}), we find that the
divergence at $x_p\to1$ cancels between the real and virtual terms, but
the divergence for $x_p<1$ due to the region $z\to1$ still remains.

\subsubsection{Factorization of divergences}

\begin{figure}[t]
  \centerline{\hfill(a)\hspace{4.5cm}\hfill
    \hfill(b)\hspace{4.5cm}\hfill}
  \vspace*{-2ex}
  \centerline{\hfill\includegraphics[79,598][218,694]{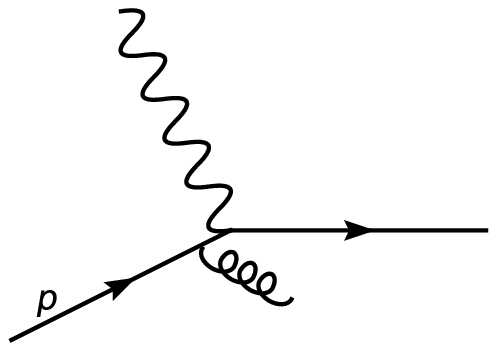}\hfill
    \hfill\includegraphics[79,598][218,694]{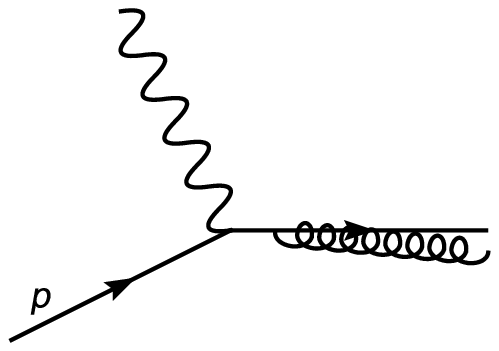}\hfill}
  \vspace*{2ex}
  \centerline{\hfill(c)\hspace{4.5cm}\hfill
    \hfill(d)\hspace{4.5cm}\hfill}
  \vspace*{-2ex}
  \centerline{\hfill\includegraphics[79,598][218,694]{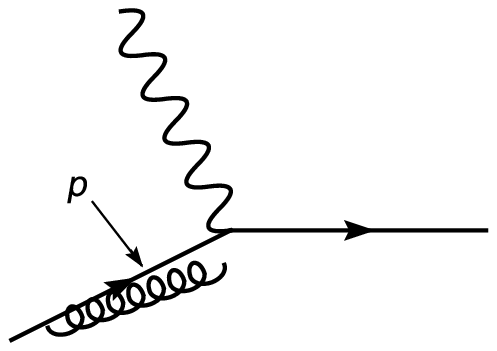}\hfill
    \hfill\includegraphics[79,598][218,694]{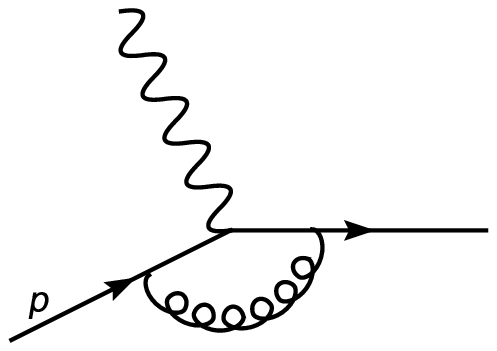}\hfill}
  \caption{Divergent contributions to DIS: (a) soft, (b) final-state
  collinear, (c) initial-state collinear, and (d) virtual.  The label $p$
  shows which momentum in each contribution is fixed by the massless
  final-state condition.}
  \label{fig:divdis}
\end{figure}

To understand why the results are still divergent even after including
the virtual terms, and what ultimately happens to the divergences, we
consider their physical origin.  Like the $e^+e^-$ annihilation case, we
have singularities from regions in which the real gluon is collinear
with either the incoming or outgoing quark, or is soft, and also from
the virtual graph, as illustrated in Fig.~\ref{fig:divdis}.
All these contributions were present in $e^+e^-$ annihilation, but there
we found that the divergences all cancelled to give a finite
contribution.  Why is the present situation different?  In fact we find
that here the magnitudes of the divergences are such as to cancel, but
that the divergences arise in different regions of the $x_p$ integral,
so are prevented from cancelling.

In the $e^+e^-$ case, we argued that the singular regions of real
emission were indistinguishable from the lowest order process, since an
infinitely soft gluon could not produce any hadrons and the jets
produced by two collinear partons were indistinguishable from a single
jet with their combined momentum.  This statement is true here for the
soft and final-state collinear contributions, but not the initial-state
contribution.  The final state of this contribution is indeed
indistinguishable from the lowest order process (it has an additional
jet collinear with the outgoing proton remnant, but this too gives a jet
and the superposition of the two is indistinguishable from the proton
remnant in the lowest order process).  However, because we have used the
parton model, we must convolute the partonic cross sections over an
arbitrary (measured from experiment) probability distribution function,
processes with different incoming momenta are effectively
distinguishable.  In all the singular regions, the final state of the
process is massless, and this fact fixes the incoming momentum (to the
value $Q/2x$ in the Breit frame), but in the initial-state singular
process it is the internal line whose momentum gets fixed, as indicated
in Fig.~\ref{fig:divdis}.  Thus the incoming momentum in (c) is larger
than in the other cases and its divergence, at $\eta>x$, cannot cancel
the others, at $\eta=x$.

As I mentioned earlier, these divergences come from the virtuality of
the internal particle vanishing and hence the propagator diverging.
Using the uncertainty principle, vanishingly small virtuality
corresponds to arbitrarily long time-scales.  This seems to be in direct
contradiction with the assumption underlying the parton model, that the
virtual photon takes an extremely rapid snapshot of the proton.

The problem is actually one of overcounting.  We first introduced the
pdfs, which are supposed to contain all information about the internal
structure of the proton.  Presumably this internal structure is the
result of QCD interactions.  We then tried to calculate the QCD
corrections to the quark scattering cross sections, integrating over all
final states, so all energy-scales.  But these QCD corrections should
somehow already be included in the internal dynamics of the proton.

To resolve this overcounting, we have to separate (or `factorize') the
different types of physics at different energy scales.  Like in our
discussion of renormalization, I will first try to give the physical
picture in terms of a cutoff, before returning later to describe how
factorization works in practice in dimensional regularization.  We
introduce the factorization scale $\mu$, and call all physics at scales
below $\mu$ part of the hadron wave function, and lump it into the
parton distribution functions, and call all physics at scales above
$\mu$ part of the partonic scattering cross section (or `coefficient
function').

Therefore we do in fact have a transverse momentum cutoff in the $eq\to
eqg$ process and the form of (\ref{Pxp}) is correct.

Since physics at scales below $\mu$ is included in the pdfs and physics
above is not, the pdfs themselves must become $\mu$-dependent.  We
therefore have
\be
  \label{F2NLO}
  F_2(x,Q^2) = \sum_qe_q^2\int_x^1 dx_p\frac{x}{x_p}
  f_q\left(\frac{x}{x_p},\mu^2\right)
  \left\{
    \delta(1-x_p)
    +\frac{\as}{2\pi}\left(P(x_p)\log\frac{Q^2}{\mu^2}
    +R(x_p)\right)
    +{\cal O}(\as^2)
  \right\},
\ee
where the function $R(x_p)$ is not necessarily the same one as earlier,
as the virtual contributions could have introduced some additional
finite terms.

Note that the structure functions are now $Q^2$-dependent, violating
Bjorken scaling.  However, they also appear to be $\mu^2$ dependent,
which should worry us: $\mu$ was introduced in a completely ad hoc
theoretical way: it simply separates physical processes into two parts
that are dealt with in different ways, and the final result, which is
the sum of the two parts, should not depend on where the separation was
made.  We return to discuss this point in more detail after calculating
the $\mu^2$-dependence of the pdfs.

It is important to emphasize that, although we have derived these
formulae for the higher order corrections to DIS, the leading
logarithmic behaviour is universal.  In particular, for any
quark-induced process with a hard scale $Q$, we expect a hadronic cross
section of the form
\be
  \sigma_h(p_h) = \sum_q \int d\eta\,
  f_q\left(\eta,\mu^2\right)
  \left\{
    \sigma_q(\eta p_h)
    +\frac{\as}{2\pi}\log\frac{Q^2}{\mu^2}
    \int dz\,P(z)\,\sigma_q(z\eta p_h)
  \right\},
\ee
where $\sigma_q(p)$ is the partonic cross section for a quark of flavour
$q$ and momentum $p$.

\subsubsection{DGLAP evolution equation}

Although the pdfs are fundamentally non-perturbative and cannot be
predicted from first principles at present, physics at scales close to
$\mu^2$ can be described perturbatively.  We can therefore calculate the
$\mu^2$-dependence of the pdfs so that, given their value at some
starting scale $\mu_0$, for example from experimental measurements, we
can calculate their values at all higher scales $\mu$.

To do this, we use the fact just noted, that physical cross sections
should not depend on $\mu^2$.  Therefore we should have
\be
  \mu^2\frac{dF_2(x,Q^2)}{d\mu^2} = 0,
\ee
or at least, since we are working at ${\cal O}(\as)$,
\be
  \mu^2\frac{dF_2(x,Q^2)}{d\mu^2} = {\cal O}(\as^2).
\ee
Applying this to (\ref{F2NLO}), we obtain
\be
  \label{DGLAP}
  \mu^2\frac{d}{d\mu^2}
  f_q\left(x,\mu^2\right)
  =
  \frac{\as}{2\pi}
  \int_x^1 \frac{dx_p}{x_p}
  f_q\left(\frac{x}{x_p},\mu^2\right)
    P(x_p)
  +{\cal O}(\as^2).
\ee
Equation (\ref{DGLAP}) is called the
Dokshitzer--Gribov--Lipatov--Altarelli--Parisi (or DGLAP, or GLAP, or
Altarelli--Parisi for short) evolution equation.  Note that the rate of
change of the pdf at some $x$ value depends on its value at all higher
$x$s.

To understand its physical content, it is useful to rewrite the
splitting function,
\be
  P(x) = C_F\left[\frac{1+x^2}{(1-x)_+}+\frac32\delta(1-x)\right]
  = C_F\left(\frac{1+x^2}{1-x_{\phantom{p}}}\right)_+,
\ee
to give
\be
  \mu^2\frac{d}{d\mu^2}
  f_q\left(x,\mu^2\right)
  =
  C_F\frac{\as}{2\pi}
  \int_x^1 \frac{dx_p}{x_p}
  f_q\left(\frac{x}{x_p},\mu^2\right)
  \frac{1+x_p^2}{1-x_p}
  -
  C_F\frac{\as}{2\pi}
  f_q\left(x,\mu^2\right)
  \int_0^1 dx_p
  \frac{1+x_p^2}{1-x_p}.
\ee
The first term represents the fact that the pdf at a given $x$ value is
increased by quarks with higher $x$'s reducing their momentum fractions
by radiating gluons.  The second term represents the fact that it is
decreased by the quarks at that $x$ reducing their momentum fractions by
radiating gluons.  Each contribution is divergent due to emission with
$x_p\to1$, i.e., infinitely soft gluon emission, involving an infinitely
small change in~$x$.  However the two divergences exactly cancel because
the number of quarks being lost to this $x$ value by infinitely soft
gluon emission is equal to the number being gained.

The DGLAP equation is most easily solved in moment space.  For any
function $f(x)$, we define
\be
  f_N = \int_0^1 dx\,x^{N-1}\,f(x),
\ee
the Mellin transform.  Taking moments of both sides of (\ref{DGLAP}), we
obtain
\bee
  \mu^2\frac{d}{d\mu^2}
  f_{qN}\left(\mu^2\right)
  &=&
  \frac{\as}{2\pi}
  \int_0^1 dx\,x^{N-1}\,
  \int_x^1 \frac{dx_p}{x_p}
  f_q\left(\frac{x}{x_p},\mu^2\right)
    P(x_p)+{\cal O}(\as^2)
  \\
  &=&
  \label{DGLAPnspace}
  \frac{\as}{2\pi}
  P_N
  f_{qN}(\mu^2).
\eee
It is common to introduce the notation
\be
  \gamma_N(\as) = \frac{\as}{2\pi}P_N+{\cal O}(\as^2),
\ee
where $\gamma_N$ is known as the anomalous dimension.  If we assume that
the coupling $\as$ is fixed, we can easily solve (\ref{DGLAPnspace})
with the boundary condition of given values for $f_{qN}$ at some fixed
scale $\mu_0$,
\be
  f_{qN}(\mu^2) =
  f_{qN}(\mu_0^2)\left(\frac{\mu^2}{\mu_0^2}\right)^{\gamma_N(\as)}.
\ee

However, as we have seen, the renormalization of QCD means that the
coupling constant becomes scale dependent, $\as(\mu^2)$, according to
renormalization group equation
\be
  \label{RGE}
  \mu^2\frac{d}{d\mu^2}\as(\mu^2) = \beta(\as(\mu^2)) =
  -\frac{\beta_0}{2\pi}\as^2(\mu^2) + {\cal O}(\as^3).
\ee
Inserting the solution of the running coupling, Eq.~(\ref{asmu}), into
(\ref{DGLAPnspace}), we obtain
\be
  f_{qN}(\mu^2) = f_{qN}(\mu_0^2)
  \left(\frac{\as(\mu_0)}{\as(\mu)}\right)^{\frac{P_N}{\beta_0}}.
\ee

Having the solution for $f_q$ in moment $N$-space, we have to convert it
back to $x$-space.  This is done by the Inverse Mellin Transform, where
$f_{qN}$ is continued to the complex plane,
\be
  f_q(x) = \frac1{2\pi i}\int_CdN\,f_{qN}\,x^{-N},
\ee
where the contour $C$ runs parallel to the imaginary axis to the right
of all poles.  Because of the complexity of this process, the DGLAP
equation is often solved simply by `brute force' numerical solution of
(\ref{DGLAP}).

Beyond ${\cal O}(\as)$ the general structure of (\ref{DGLAPnspace}) and
(\ref{RGE}) remains unchanged: the anomalous dimension and $\beta$
function simply become power series in $\as$.

\subsubsection{Scheme/scale dependence}

Factorization, as introduced above, may seem pretty arbitrary.  However
it can be proved to all orders in perturbation theory.  The most
convenient way to do this is to use, instead of the transverse momentum
cutoff we used above, dimensional regularization.  When we calculate the
NLO cross section in $d$ dimensions, the divergence shows up as a
pole,~$1/\epsilon$.  The coefficient multiplying this pole turns out to
be the same splitting function we encountered earlier.

In $d$ dimensions, we obtain for the structure function up
to~${\cal O}(\as)$,
\be
  \label{regularization}
  \hspace*{-5mm}
  F_2(x,Q^2) = \sum_qe_q^2\int_x^1 dx_p\frac{x}{x_p}
  \bar f_q\left(\frac{x}{x_p}\right)
  \left\{
    \delta(1-x_p)
    +\frac{\as}{2\pi}\left(
      \left(\frac{4\pi\mu^2}{Q^2}\right)^\epsilon\frac{-1}\epsilon P(x_p)
    +R(x_p)\right)
    +{\cal O}(\epsilon)
  \right\},
  \hspace*{-5mm}
\ee
where $\mu$ is the scale introduced to make the coupling constant
dimensionless.  Note that I have sneakily added a bar to $f_q$ and that
it is scale independent.  $\bar f_q$ is known as the bare pdf.  We now
note that the distribution functions themselves are not physical
observables, only their convolution with coefficient functions is.  I can
therefore define a modified set of distribution functions as follows:
\be
  \label{subtraction}
  x\,\bar f_q(x) \equiv \int_x^1 dx_p\frac{x}{x_p}
  f_q\left(\frac{x}{x_p},\mu_F^2\right)
  \left\{
    \delta(1-x_p)
    -\frac{\as}{2\pi}\left(
      \left(\frac{4\pi\mu^2}{\mu_F^2}\right)^\epsilon
      \frac{-1}\epsilon P(x_p)
    +K(x_p)\right)
  \right\},
\ee
where $\mu_F$ is the (completely arbitrary again) factorization scale,
and $K(x_p)$ is a completely arbitrary finite function to be discussed
shortly.  (To fit in with the standard notation, I should really
multiply all occurrences of $\as$ by $1/\Gamma(1-\epsilon) =
1-\gamma_E\epsilon+{\cal O}(\epsilon^2)$, but this will merely change
the values of $R(x_p)$ and $K(x_p)$ which I do not specify anyway.)
Combining (\ref{regularization}) and (\ref{subtraction}), we end up with
\begin{multline}
  F_2(x,Q^2) = \sum_qe_q^2
  \int_x^1 dx_p\frac{x}{x_p}
  f_q\left(\frac{x}{x_p},\mu_F^2\right)
  \left\{
    \delta(1-x_p) \vphantom{\frac{Q^2}{\mu_F^2}}\right.\\
    \left.
    +\frac{\as}{2\pi}\left(
      P(x_p)\log\frac{Q^2}{\mu_F^2}
    +R(x_p)-K(x_p)\right)
    +{\cal O}(\as^2)
  \right\}.
\end{multline}  
Note that this has the identical form to (\ref{F2NLO}), except for the
finite function.  It is clear from (\ref{subtraction}) that
$f_q(x,\mu_F^2)$ depends on the function $K(x_p)$.  It therefore seems
like we have no predictive power: the pdf and coefficient function each
depend on the completely arbitrary function $K(x_p)$ and the completely
arbitrary scale $\mu_F$ (note that all dependence on $\mu$ has again
completely cancelled.  As I said in the context of renormalization, many
textbooks simply set it equal $\mu$ right from the start, but I consider
this slightly confusing as they have quite different physical meaning.
Having performed this manoeuvre, I henceforth drop the subscript
${}_F$).  However, the factorization theorem proves, firstly that for any
physical quantity, all dependence on $K(x_p)$ and $\mu$ will cancel and
secondly that the scheme- and scale-dependent pdfs, $f_q(x,\mu^2)$ are
universal (i.e., process-independent).

Two schemes are in common use, the $\overline{\mathrm{MS}}$ scheme in
which $K(x_p)$ is zero, and the DIS scheme in which $K(x_p)=R(x_p)$,
i.e.~in which for $\mu=Q$ the parton model result is exact.

To understand the physical content of the scheme-dependence, it is
worth while going back to the case with a cutoff.  If, instead of
a cut on transverse momentum we had used a cut on the virtuality of the
internal quark line to separate the pdf from the coefficient function,
we would have got exactly the same form as (\ref{F2NLO}) except that
$R(x_p)$ would have been a different function.  In particular, it would
differ by a $\log[(1-x_p)/x_p]$ term, together with some non-logarithmic
terms.  In fact, all logarithmic terms turn out to be the same with a
$p_t$ cutoff as in the $\overline{\mathrm{MS}}$ scheme, so for many
purposes the two can be considered equivalent.

Although dependence on the scheme and scale must cancel in physical
quantities, it is only guaranteed to do so after calculating to infinite
orders of perturbation theory.  At any finite order there can be some
residual dependence.  We must therefore have a procedure for choosing a
value of $\mu$.  Essentially the identical discussion we had for the
renormalization scale choice applies here.  One can show that a structure
like (\ref{F2NLO}) continues to all orders of perturbation theory and
that for every power of $\as$, one gets a power of $\log Q^2/\mu^2$.
Thus every order of perturbation theory contains terms like
$\as^n\log^mQ^2/\mu^2$, $m\le n$.  It is clear that if $\mu$ is a long
way from $Q$, the log can be large enough to compensate the smallness of
$\as$ and the perturbative series will not converge quickly.  One should
therefore choose $\mu$ `not too far' from $Q$.

It is worth mentioning that one can set up DGLAP evolution equations for
the $Q^2$-dependence of the structure functions, $F_2$ and $F_L$,
themselves.  These are then automatically scheme- and scale-independent
even at finite orders of perturbation theory.  This is sometimes known
as the scheme-independent scheme.

\subsubsection{Initial-state gluons}

\begin{figure}[t]
  \centerline{\hfill\includegraphics[141,628][248,661]{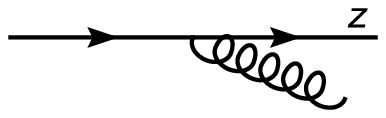}\hfill
    \hfill\includegraphics[141,628][248,661]{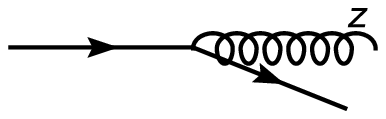}\hfill}
  \centerline{\hfill\makebox[0pt]{$\displaystyle
  P_{qq}(x) =
  C_F\Biggl[\frac{1+x^2}{(1-x)_+}+\frac32\delta(1-x)\Biggr]
  $}\hfill
    \hfill\makebox[0pt]{$\displaystyle
  P_{qg}(x) = T_R\Biggl[x^2+(1-x)^2\Biggr]
  $}\hfill}
  \vspace*{4ex}
  \centerline{\hfill\includegraphics[141,628][248,661]{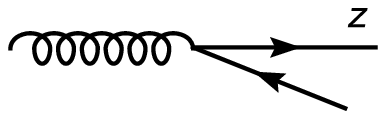}\hfill
    \hfill\includegraphics[141,628][248,661]{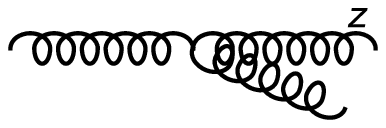}\hfill}
  \centerline{\hfill\makebox[0pt]{$\displaystyle
  P_{gq}(x) = C_F\Biggl[\frac{1+(1-x)^2}x\Biggr]
  $}\hfill
    \hfill\makebox[0pt]{$\displaystyle
  P_{gg}(x) = C_A\Biggl[\frac{2x}{(1-x)_+} + 2\frac{1-x}x + 2x(1-x)\Biggr]
  $}\hfill}
  \centerline{\hfill\hfill\hfill\makebox[0pt]{$\displaystyle
  + \beta_0\delta(1-x)
  $}\hfill}
  \caption{The four DGLAP splitting functions of QCD}
  \label{fig:Pij}
\end{figure}

As mentioned right at the start of this section, we also obtain ${\cal
  O}(\as)$ corrections from the process $eg \to eq\bar q$.  Most of what
we said above carries over in a straightforward way.  Although there is
no soft singularity or virtual term to cancel it, there is a collinear
singularity.  This corresponds to a two-step process in which a gluon
splits to a $q$--$\bar q$ pair, one of which interacts with the photon.
The singularity again corresponds to the virtuality of the internal
quark line going to zero.  This singularity can again be absorbed into a
factorized universal pdf for the gluon.  We end up with an additional
contribution to the structure function of
\be
  \hspace*{-5mm}
  F_2(x,Q^2) = \sum_qe_q^2
  \int_x^1 dx_p\frac{x}{x_p}
  f_g\left(\frac{x}{x_p},\mu^2\right)
  \left\{
    \frac{\as}{2\pi}\left(
      P_{qg}(x_p)\log\frac{Q^2}{\mu^2}
      +R_g(x_p)-K_{qg}(x_p)\right)
    +{\cal O}(\as^2)
  \right\},
  \hspace*{-5mm}
\ee
where the sum over $q$ is over all `light' flavours.  We now have four
different types of splitting function, illustrated in
Fig.~\ref{fig:Pij}.
The DGLAP equation now becomes a set of coupled equations:
\be
  \mu^2\frac{d}{d\mu^2}
  f_a\left(x,\mu^2\right)
  =
  \sum_b
  \frac{\as}{2\pi}
  \int_x^1 \frac{dx_p}{x_p}
  f_b\left(\frac{x}{x_p},\mu^2\right)
    P_{ab}(x_p)
  +{\cal O}(\as^2).
\ee
In moment space, this can be conveniently written as a matrix equation
(in general of $(2N_f\!+\!1)\times(2N_f\!+\!1)$ matrices, but for
simplicity we show the case of only one flavour of quark):
\be
  \label{zeroes}
  \mu^2\frac{d}{d\mu^2}
  \left(\begin{array}{c}f_{qN}\\f_{\bar qN}\\f_{gN}\end{array}\right)
  =
  \left(\begin{array}{ccc}
      \gamma_{qqN}(\as(\mu))&0&\gamma_{qgN}(\as(\mu))\\
      0&\gamma_{qqN}(\as(\mu))&\gamma_{qgN}(\as(\mu))\\
      \gamma_{gqN}(\as(\mu))&\gamma_{gqN}(\as(\mu))&\gamma_{ggN}(\as(\mu))
    \end{array}\right)
  \left(\begin{array}{c}f_{qN}\\f_{\bar qN}\\f_{gN}\end{array}\right).
\ee
Exactly the same solution is obtained, but in matrix notation,
\be
  \left(\begin{array}{c}
      f_{qN}(\mu^2)\\f_{\bar qN}(\mu^2)\\f_{gN}(\mu^2)
    \end{array}\right)
  = \exp\int_{\mu_0^2}^{\mu^2} \frac{d\mu'^2}{\mu'^2}
  \left(\begin{array}{ccc}
      \gamma_{qqN}(\as(\mu'))&0&\gamma_{qgN}(\as(\mu'))\\
      0&\gamma_{qqN}(\as(\mu'))&\gamma_{qgN}(\as(\mu'))\\
      \gamma_{gqN}(\as(\mu'))&\gamma_{gqN}(\as(\mu'))&\gamma_{ggN}(\as(\mu'))
    \end{array}\right)
  \left(\begin{array}{c}
      f_{qN}(\mu_0^2)\\f_{\bar qN}(\mu_0^2)\\f_{gN}(\mu_0^2)
    \end{array}\right).
\ee
This is even more troublesome to do by the Inverse Mellin Transform, so
the full set of DGLAP equations is almost always solved numerically.

Note that at higher orders of perturbation theory, even the zero entries
in (\ref{zeroes}) become non-zero, as do contributions
like~$P_{qq'}(x)$.

\subsubsection{Violation of Bjorken scaling}

\begin{figure}[t]
  \centerline{\scalebox{0.5}{\includegraphics{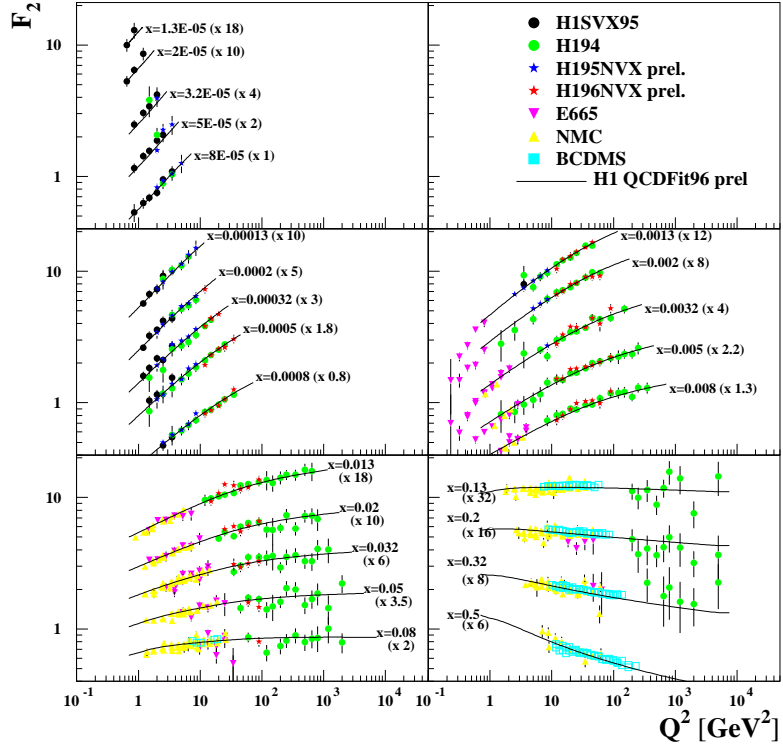}}}
  \caption{Fit to the $F_2$ data over a wide range of $Q^2$ values,
  exhibiting violation of Bjorken scaling}
  \label{fig:bjviolation}
\end{figure}

As we already noted, the factorization of initial-state singularities
introduces a logarithmic $Q^2$ dependence into the structure functions
and therefore a slow violation of Bjorken scaling.  There is a close
analogy with the renormalization of one-scale cross sections, where the
energy-dependence was entirely due to the quantum corrections.  Although
the pdfs at some low scale are entirely non-perturbative and must be fit
to data, the scale-dependence is entirely predicted by QCD and provides
a stringent test over a wide range of energy scales.  The result is
impressive, see Fig.~\ref{fig:bjviolation}.

\subsection{Summary}

NLO calculations are hard!  This is mainly because the real and virtual
corrections are each divergent and must be regularized in some
self-consistent way, for example with dimensional regularization.
Unlike the ultraviolet divergences, which are isolated in well-localized
pieces of the loop calculation and can effectively be removed by a
redefinition of the Feynman rules, these divergences arise in different
partonic contributions to physical observables.  They must therefore be
kept explicit until the very end of the calculation when all the
partonic contributions are combined.  Only then, provided our observable
is infrared safe, will the real and virtual divergences cancel to yield
a finite result.

Processes with incoming partons have extra divergences, arising from a
miscancellation of the initial-state-collinear real and virtual
contributions, which appear at different points in the integral over
incoming momentum fraction.  (It is worth mentioning that the same
argument applies to the final-state distributions of identified hadrons,
for example the momentum distribution of pions produced in $e^+e^-$
annihilation.)  These divergences have to be factorized into the
non-perturbative, but universal, parton distribution functions at some
factorization scale $\mu_F$.  This extra scale in the structure
functions allows them to be $Q^2$-dependent.  This $Q^2$-dependence is
entirely driven by the $\mu_F^2$-dependence of the parton distribution
functions, which is predicted by the DGLAP evolution equations.  Thus
structure function data over a wide range of $Q^2$ provide a stringent
test of perturbative QCD.

\section{Summary}
\vspace*{-0.3ex}

In this short course on QCD phenomenology, I have resisted the
temptation to review the many important tests and studies of QCD that
have been made over the years and have instead tried to concentrate on
the key ideas that underpin them.  These are:
\begin{itemize}
\item The gauge invariance of the theory, which allows us to write down
  the Lagrangian and which predicts some of the most important features
  of the theory: the universality of the coupling constant and the
  self-coupling of gluons, which ultimately leads to the negative
  $\beta$ function and hence to asymptotic freedom at high energies and
  strong interactions at low energies.
\item Renormalization and decoupling, which allow us to make predictive
  calculations at finite energy, without knowing the full structure of
  the theory to arbitrarily high energy and without the introduction of
  arbitrarily many input parameters.  Renormalization is related to the
  quantum structure of the theory and introduces a dimensionful scale
  into even the scaleless Lagrangian of massless QCD, giving rise to
  energy-dependence of one-scale observables that would be
  energy-independent in the classical theory.
\item Factorization and evolution, which allow us to use perturbation
  theory to calculate the interactions of hadrons, since all the
  non-perturbative physics gets factorized, into universal functions that
  can be measured in one process, like DIS, and then used to predict the
  cross sections for any other process.  Again, this introduces a scale
  dependence into the parton model so that the structure functions of
  DIS, and other one-scale observables such as the Drell--Yan cross
  section, become scale dependent.
\item Infrared safety, which ensure that the infrared singularities
  associated with soft and collinear emission cancel between real and
  virtual contributions, allowing the perturbative calculation of jet
  cross sections, without a detailed understanding of the mechanism by
  which partons become jets.
\end{itemize}
Together, these allow us to make sense of QCD, without having to solve
the theory at all possible scales: unknown or uncalculable high- and
low-energy effects can be renormalized, factorized and cancelled away.
After all this, it is remarkable that most QCD phenomenology can be
understood at least qualitatively from leading order perturbation theory
with the one-loop renormalization group and DGLAP evolution equations.
Higher order corrections, while essential for quantitative analysis, do
not change this simple picture dramatically.

\vspace*{-2.3ex}
\section*{Acknowledgements}
\vspace*{-0.3ex}

It is a pleasure to acknowledge the organizers, lecturers, tutors and
students of the Latin American School of High Energy Physics at Recinto
Quirama for making giving these lectures such an enjoyable experience.


\vspace*{-2.3ex}
\begin{thebibliography}{99}
\vspace*{-0.3ex}
\bibitem{ESW}
  R.K.~Ellis, W.J.~Stirling, and B.R.~Webber, {\it QCD and Collider
    Physics}, Cambridge Monographs on Particle Physics, Nuclear Physics
  and Cosmology, Volume 8 (Cambridge University Press, 1996).
\bibitem{Peskin}
  M.E.~Peskin and D.V.~Schroeder, {\it An Introduction to Quantum Field
    Theory} (Addison-Wesley, 1995).
\bibitem{Gockeler:2005rv}
  M.~Gockeler, R.~Horsley, A.~C.~Irving, D.~Pleiter, P.~E.~L.~Rakow, G.~Schierholz, and H.~Stuben,
  {\it A determination of the Lambda parameter from full lattice QCD},
  Phys.\ Rev.\  D {\bf 73} (2006) 014513
  [arXiv:hep-ph/0502212].
\bibitem{Davier:2002dy}
  M.~Davier, S.~Eidelman, A.~H\"ocker, and Z.~Zhang,
  {\it Confronting spectral functions from e+ e- annihilation and tau decays:
  Consequences for the muon magnetic moment},
  Eur.\ Phys.\ J.\  C {\bf 27} (2003) 497
  [arXiv:hep-ph/0208177].
\bibitem{Davier:2008sk}
  M.~Davier, S.~Descotes-Genon, A.~H\"ocker, B.~Malaescu, and Z.~Zhang,
  {\it The determination of $\alpha_s$ from $\tau$ decays revisited},
  Eur.\ Phys.\ J.\  C {\bf 56} (2008) 305
  [arXiv:0803.0979 [hep-ph]].
\bibitem{Bethke:2006ac}
  S.~Bethke,
  {\it Experimental tests of asymptotic freedom},
  Prog.\ Part.\ Nucl.\ Phys.\  {\bf 58} (2007) 351
  [arXiv:hep-ex/0606035].
\end{thebibliography}
\end{document}